\documentclass[12pt]{cernrep}
\usepackage{graphicx}
\usepackage{here}
\usepackage{epsfig}

\begin{document}
 \vspace*{2truecm}

 
\begin{titlepage}
\begin{flushright}
ETH-TH/2000-06\\
April 1999
\end{flushright}
\begin{center}
\vspace*{2.5cm}
{
\large \bf $\quad$ BREAD AND BUTTER STANDARD MODEL}
 
\vskip 0.5cm
{\large Z. Kunszt} \\
\vskip 0.4cm
{\it Theoretical Physics, ETH Z\"urich, Switzerland} \

\vspace*{1.2truecm}
\begin{center}
 \begin{minipage}{0.5\linewidth}

{\centerline{\bf\quad Contents:\\}}
\vspace*{0.5truecm}
\begin{itemize}
\item[1.]
Structure of the Standard Model
\begin{itemize}
\vspace*{0.2truecm}
 \item[ 1.1]
 Basic QCD
  \item[1.2]
 Basic Electroweak Theory\\
\end{itemize}
\item[2.] 
Precision calculations
\begin{itemize}
\vspace*{0.2truecm}
   \item[ 2.1] Testing QCD
    \item[ 2.2] Testing the Electroweak Theory\\
\end{itemize}
\item[3.] Higgs sector, Higgs search
\begin{itemize}
\vspace*{0.2truecm}
\item[3.1]Difficulties with the Higgs sector
\item[3.2]Higgs search at LHC
\end{itemize}
\end{itemize}

 \end{minipage}

\end{center}

\end{center}

\vfill
\noindent\hrule width 3.6in\hfil\break
{\small 
Lectures given at International School of Subnuclear Physics: 37th Course: Basics and Highlights of Fundamental Physics, Erice, Italy, 29 Aug - 7
Sep 1999.}
\hfil\break

\end{titlepage}
\setcounter{footnote}{0}

\def\eg{{\it e.g.}}
\def\etc{{\it etc.}}
\def\etal{{\it et al.}}
\def\ie{{\it i.e.}}
\def\dash{\hbox{---}}
\def\abs#1{\left| #1\right|}   
\def\VEV#1{\left\langle #1\right\rangle}
\def\lsim{{\buildrel < \over\sim}}
\def\gsim{{\buildrel > \over\sim}}
\def\to{\rightarrow}

\def\pbi{~{\rm pb}^{-1}}
\def\fbi{~{\rm fb}^{-1}}
\def\fb{~{\rm fb}}
\def\pb{~{\rm pb}}
\def\ev{\,{\rm eV}}
\def\mev{\,{\rm MeV}}
\def\gev{\,{\rm GeV}}
\def\tev{\,{\rm TeV}}
\newcommand{\TeV}{{\rm TeV}}
\newcommand{\GeV}{{\rm GeV}}
\newcommand{\MeV}{{\rm MeV}}
\def\MSbar{\overline{\rm MS}}
\newcommand\msbar{\ifmmode{\overline{\rm MS}}\else $\overline{\rm MS}$\ \fi}
\def\als{\alpha_s}
\newcommand{\as}{{\ifmmode \alpha_S \else $\alpha_S$ \fi}}
\newcommand\asopi{ \frac{\as}{\pi} }
\newcommand\aopi{ \frac{\alpha}{\pi} }
\newcommand{\asotwopi}{ \frac{\as}{2\pi} }
\newcommand{\LMSB}{\ifmmode \Lambda_{\overline{\rm MS}} \else
          $\Lambda_{\overline{\rm MS}}$ \fi}
\newcommand{\aem}{\alpha_{\rm em}}
\newcommand\sinthw{\sin^2(\theta_{\rm W})}
\newcommand\epem{\ifmmode e^+e^- \else $e^+e^-$ \fi}
\newcommand\mupmum{ \mu^+\mu^- }
\newcommand{\ep}{\epsilon}
\newcommand\epbar{\overline\epsilon}
\newcommand\nf{\alwaysmath{{n_{\rm f}}}}

\def\ttbs{\char'134}
\def\AmS{{\protect\the\textfont2
  A\kern-.1667em\lower.5ex\hbox{M}\kern-.125emS}}

\def\bea{\begin{eqnarray}}
\def\be{\begin{equation}}
\def\bear{\begin{eqnarray*}}
\def\eea{\end{eqnarray}}
\def\ee{\end{equation}}
\def\eear{\end{eqnarray*}}
\catcode`@=11
\def\Biggg#1{\hbox{$\left#1\vbox to 22.5\p@{}\right.\n@space$}}
\catcode`@=12
\def\mathswitchr#1{\relax\ifmmode{\mathrm{#1}}\else$\mathrm{#1}$\fi}
\newcommand{\PB}{\mathswitchr B}
\newcommand{\Pn}{\mathswitchr n}
\newcommand{\Pp}{\mathswitchr p}
\newcommand{\PW}{\mathswitchr W}
\newcommand{\PZ}{\mathswitchr Z}
\newcommand{\Pg}{\mathswitchr g}
\newcommand{\PH}{\mathswitchr H}
\newcommand{\Pe}{\mathswitchr e}
\newcommand{\Pmu}{\mathswitchr \mu}
\newcommand{\Ptau}{\mathswitchr \tau}
\newcommand{\Pne}{\mathswitch \nu_{\mathrm{e}}}
\newcommand{\Pane}{\mathswitch \bar\nu_{\mathrm{e}}}
\newcommand{\Pnmu}{\mathswitch \nu_\mu}
\newcommand{\Pntau}{\mathswitch \nu_\tau}
\newcommand{\Pf}{f}
\newcommand{\Ph}{\mathswitchr h}
\newcommand{\Pl}{l}
\newcommand{\Pd}{\mathswitchr d}
\newcommand{\Pu}{\mathswitchr u}
\newcommand{\Ps}{\mathswitchr s}
\newcommand{\Pb}{\mathswitchr b}
\newcommand{\Pc}{\mathswitchr c}
\newcommand{\Pt}{\mathswitchr t}
\newcommand{\Pq}{\mathswitchr q}
\newcommand{\Pep}{\mathswitchr {e^+}}
\newcommand{\Pem}{\mathswitchr {e^-}}
\newcommand{\Pepm}{\mathswitchr {e^\pm}}
\newcommand{\Pmum}{\mathswitchr {\mu^-}}
\newcommand{\Ptaum}{\mathswitchr {\tau^-}}
\newcommand{\PWp}{\mathswitchr {W^+}}
\newcommand{\PWm}{\mathswitchr {W^-}}
\newcommand{\PWpm}{\mathswitchr {W^\pm}}

\def\mathswitch#1{\relax\ifmmode#1\else$#1$\fi}
\def\mt{m_t}
\def\mb{m_b}
\def\mz{m_Z}
\def\mw{m_W}
\newcommand{\MB}{\mathswitch {M_\PB}}
\newcommand{\MZ}{\mathswitch {M_\PZ}}
\newcommand{\MW}{\mathswitch {M_\PW}}
\newcommand{\Mf}{\mathswitch {m_\Pf}}
\newcommand{\Ml}{\mathswitch {m_\Pl}}
\newcommand{\Mq}{\mathswitch {m_\Pq}}
\newcommand{\MV}{\mathswitch {M_\PV}}
\newcommand{\MWpm}{\mathswitch {M_\PWpm}}
\newcommand{\MWO}{\mathswitch {M_\PWO}}
\newcommand{\MA}{\mathswitch {\lambda}}
\newcommand{\MH}{\mathswitch {M_\PH}}
\newcommand{\Me}{\mathswitch {m_\Pe}}
\newcommand{\Mmy}{\mathswitch {m_\mu}}
\newcommand{\Mpi}{\mathswitch {m_\pi}}
\newcommand{\Mta}{\mathswitch {m_\tau}}
\newcommand{\Md}{\mathswitch {m_\Pd}}
\newcommand{\Mu}{\mathswitch {m_\Pu}}
\newcommand{\Ms}{\mathswitch {m_\Ps}}
\newcommand{\Mc}{\mathswitch {m_\Pc}}
\newcommand{\Mb}{\mathswitch {m_\Pb}}
\newcommand{\Mt}{\mathswitch {m_\Pt}}


\vspace*{1.6truecm}

{\section{ STRUCTURE OF THE STANDARD MODEL}



The Standard Model is our  theory for
the quantitative descriptions of all interactions of fundamental
particles except quantum gravity effects.
 It is highly successful:
all  measurements are in agreement with the Standard Model
predictions.

The Standard Model is a renormalizable relativistic quantum field theory
based on  non-Abelian gauge symmetry~\cite{'tHooft:1994gh}
 of the gauge group
 ${ SU(3)_{C} \times SU(2)_{L}\times U(1)_Y}$.
 It has two sectors: Quantum Chromodynamics (QCD)
 and the
Electroweak Theory (EW). QCD 
is a vector gauge theory which 
describes the $SU(3)_C$ color interactions of quarks and gluons.
 It has  rich dynamical structure such as  chiral
symmetry breaking, asymptotic freedom, quark confinement, topologically
non-trivial configurations (monopoles, instantons).
The Electroweak Theory (EW) describes the electromagnetic and weak  
 interactions of the quarks and leptons as a  
 chiral non-Abelian
 isospin and an Abelian hypercharge gauge symmetry
 $SU(2)_L\times U(1)_Y$. As a result of the 
Higgs mechanism, 
 the gauge bosons 
$W^{\pm},Z$ become massive while the photon remains massless.
The true dynamics behind the Higgs mechanism 
 is not yet
known. The simple one doublet Higgs sector  predicts the existence
of a single Higgs boson with well defined properties and its experimental
search has first priority. 
  Quarks carry both  color and  electroweak charges.
Quarks and leptons  
cooperate  to cancel the weak gauge anomalies. 
The Lagrangian of the  Standard Model has important accidental global
symmetries leading to 
baryon number and individual lepton number conservation in
all orders of perturbation theory without implying absolute conservation
of these quantum numbers.

\subsection{Basic QCD}

In the sixties and early seventies 
an exciting series of   beautiful  experiments 
 with many puzzling and
unexpected results 
have lead to the discovery of QCD. 

\subsubsection{Quarks, flavor, color}
Spin 1/2 quarks as elementary constituents
of strongly interacting hadrons have been invented by 
Gell-Mann and Zweig in 1964 to explain
the approximate SU(3) spectral symmetry of   baryons and mesons.
They  come in three flavor (up, down and strange)
 and form the fundamental
triplet representation of the approximate SU(3) symmetry.
The bound state wave functions
of the spin 3/2 baryons decouplet composed from such objects, however,
did not follow
the Fermi-Dirac statistics. 
Color was invented
by Greenberg in 1964
to restore the 
validity of   the correct spin-statistics, requiring 
that every quark with  a given flavor  comes in three colors (red, blue,
yellow). 
The measured normalization of the
 decay rate $\Gamma(\pi^0\to 2\gamma)$ 
and of the cross-section $\sigma (\epem\to {\rm hadrons})$
 dramatically confirmed 
  this assumption. 
The low lying hadron spectrum also had a more delicate
chiral $SU(3)_L\times SU(3)_R$ symmetry that was broken spontaneously
and explicitly and was described  in terms of algebra of currents.
The nature of the color interactions was not clear.
For example,  motivated by the success of current algebra,
Gell-Mann  suggested that the underlying field theory
of strong interactions is a quark-gluon theory with
one Abelian colorless gluon. Nambu  instead assumed that the
gluons form the octet representation of the color group $SU(3)_C$.

These  qualitative physical concepts, however, 
could not yet be summarized  into  a consistent quantitative theory.
As  next  development, the quark constituent
picture got confirmed by deep inelastic electron and
neutrino experiments.
The results   have naturally been interpreted  
as the backward scattering of electrons and neutrinos on  
free pointlike  constituents of the proton (parton model).
Their quantum numbers could be extracted from the data
and it turned out that the partons are 
quarks invented to explain hadron  spectroscopy.
The final theory could not be formulated since 
 the parton model was based on the assumption 
 of having  approximately 
free point like
constituents at short distances  within the bound state wave function.
This 
was  ``not consistent with the known class of renormalizable
field theories'' (Feynman)~\cite{Feynman:1969ej}.

\subsubsection{Breakthrough by 't Hooft}
The breakthrough was achieved  by 't Hooft in 1971 by
proving that 
 non-Abelian gauge theories are renormalized: 
 a new
class of field theories have been discovered with 
strikingly new properties.
 The quantization of non-Abelian gauge  theories is far more
complex than the 
 well-known case of  Quantum Electrodynamics (QED) because of 
the self-interaction of the gauge field.
 The algebraic complexity of the Feynman rules as well as
the Ward-identities of the exact gauge symmetry made the
study very difficult. These theories, however,  were  
 considered by many people as irrelevant (non-physical)
 because the 
 gauge bosons are necessarily massless.
't~Hooft's proofs of  the renormalizability of massless  non-Abelian gauge
theories~\cite{thooft1} and of massive gauge theories~\cite{thooft2} with 
Higgs-mechanism 
opened the possibility to find the fundamental
 theory of strong interactions as well as the electroweak interactions.

\subsubsection{Towards QCD} 
The discovery of 
the renormalizability of Yang-Mills theory by 't~Hooft helped the model
builders to put  together the 
 concepts of quarks, color and flavor as the 
basic ingredients of a non-Abelian field theory 
called Quantum Chromodynamics. 
Gell-Mann \etal~\cite{FritzschGellmann,Fritzsch:1973pi}  pointed out that if  
  the previously suggested quark-gluon model of strong interactions
is modified by replacing the Abelian colorless gluon   with
a non-Abelian colored gluon sector with exact $SU(3)_C$  gauge symmetry, 
one obtains 
 better agreement with the experimentally established
 qualitative   features of strong interactions.
They have also speculated  that
in these theories quarks and gluons might be permanently confined
(quark confinement), chiral symmetry breaking could take place and
the  so called  $U(1)_A$ 
problem may be solved.  
In addition, a completely new fundamental 
property of Yang-Mills theories was
 discovered: at shorter and shorter distances the physics looks
the same but the  interaction of the particles are reduced 
\cite{thooftmarseille, grosswilczekprl,politzer}.
In contrary to the old type of field theories, Yang-Mills theories 
are well defined at short distances.
A field
theory is 
asymptotically
free if and only if it is a non-Abelian gauge theory.
In addition,  the importance of
asymptotic freedom in connection with Bjorken scaling was
also realized and the possibility to use perturbative
methods to calculate strong interaction effects has been
pointed out.
Gross and Wilczek 
 using Wilson's operator product expansion and
renormalization group method   have shown that 
 the short and long distance contributions can  be factorized
and the short distance part can be consistently described using
perturbation theory~\cite{Gross:1973ju}.
 They have  derived the parton picture  and   interpreted   
Bjorken-scaling of deep inelastic scattering as  leading order 
absorption of the virtual photon by free quarks inside the proton.
As they could evaluate  the   corrections 
in first order, the predictions got spectacularly confirmed
by a long experimental effort.
By reformulating these results   in terms of Feynman diagrams, 
the so called QCD improved parton model has been established. It  
provides us a well-defined   algorithm 
for calculating  cross-sections of   hard scattering
processes  involving  hadrons precisely.
In particular, jet, W, Z  and heavy quark  production have been predicted 
and   properties of  bound states involving heavy quarks
 could be calculated. The discovery of heavy particles offered 
new experimental possibilities to test the QCD improved parton
model predictions.

\subsubsection{Confinement}
Asymptotic freedom implies that 
at long distances the color interactions are
strong. The system condenses in some way.
Quarks and gluons 
may get permanently confined within the hadronic bound states such
that massless gauge-bosons do not appear in the particle spectrum.
 Wilson has pointed out that
quark-confinement is a direct  consequence of local gauge
symmetry in the (non-physical) strong coupling limit when 
 QCD is formulated  on a four  dimensional Euclidean lattice~\cite{wilsonconf}.
It has been suggested that  color neutralization is energetically favored
in comparison with color separation. It is generally accepted by now that 
the   mechanism of color confinement is 
due to the condensation of magnetic
monopoles{\footnote{ Magnetic monopoles appear when the 
gauge is completely 
fixed  such that the  so called Gribov ambiguity is avoided.}
  suggested by 't Hooft~\cite{thooftconf}.  The vacuum of color dynamics is 
a dual superconductor where instead of condensate of Cooper-pairs
one has the condensate of magnetic monopoles. The best formulation
of the non-perturbative domain of gluon dynamics is the lattice gauge theory
and 't Hooft's mechanism of quark confinement has been
supported  by the results of a number of numerical
 simulation work~\cite{haymaker}.

\subsubsection{Extended objects, topologically non-trivial gauge configurations }
By studying extended objects we can get information
on the non-perturbative aspects of the theory. They
are interpreted as particles with masses inversely
proportional to the coupling constant ('t Hooft-Polyakov
magnetic monopoles \cite{tHooftmagmonopole, polyakov}). Extended solutions of the classical
field equations in Euclidean space (instantons) can produce
tunneling effects with amplitude depending exponentially
on the inverse of the coupling constant~\cite{tHooft:instantons}.
 In QCD, instantons  are important   
 for   breaking  the global flavor $U(1)_A$
 symmetry and providing strong CP-violation effects ($\theta$-term).

\subsubsection{The Lagrangian of QCD~\cite{books}}
The transformation matrix of the fundamental representation of the
local $SU(3)$ gauge group is
\begin{equation}
\Omega(x)_{ab}=\left(e^{iT^A\xi^A(x)}\right)_{ab}\,, 
\quad
T^A=\frac{1}{2}\lambda^A
\end{equation}
where $\lambda^A$ denotes the $SU(3)$ Gell-Mann matrices and $\xi^A(x)$
is the group parameter   $A=1,\ldots 8$ and $a,b=1,2,3$.
The Dirac spinor of the quarks transforms like
\be
q^{\prime}_a(x)=\Omega(x)_{ab} q_b(x) 
\ee
while 
the gauge (gluon) fields transform inhomogenously
\be 
T^C G^{C}(x)=\Omega(x) T^C G^{C}(x)\Omega^{-1}+\frac{i}{g_s}
\left(\partial_{\mu}\Omega(x)\right)\Omega^{-1}(x) 
\ee
After the choice of the matter field,  renormalizability
and  gauge symmetry dictates uniquely the form of the
classical Lagrangian.
Introducing the non-Abelian field strength 
\bea
G^A_{\mu\nu}&=&\partial_{\mu} G^A_{\nu} - \partial_{\nu} G^A_{\mu}
    -g_s f^{ABC}G^B_{\mu}G^C_{\nu}\\
D_{ab}^{\mu}&=&\delta_{ab}
\partial^{\mu} + ig_S T^C_{ab}G^{\mu C}
\eea
where $f^{ABC}$ is the structure constant of the $SU(3)$ Lie algebra, one gets 
\bea
{\cal L}_{\rm classical}&=& -\frac{1}{4}G^A_{\mu\nu}G_A^{\mu\nu} + 
\sum_f \overline{q}^{f a}(x) i \gamma_{\mu} D^{\mu}_{ab} q(x)^f_b\\ \nonumber
&&-\sum_f m_f \overline{q}^{f a}(x)q(x)_{f a}\,,\quad f=u,d,\ldots,t
\eea
where the spinor label are suppressed.
Mass terms are allowed since  QCD is a vector theory:
the color properties of the
left and right handed quarks are the same.
Renormalizability and gauge invariance, however, 
allows to add an  additional term in the Lagrangian 
\be
{\cal L}_{\theta}=\frac{\theta g_s^2}{32\pi^2}G^A_{\mu\nu}\tilde{G}_A^{\mu\nu}
\,,\quad   \tilde{G}_A^{\mu\nu}=\frac{1}{2}\epsilon^{\mu\nu\lambda\rho}
G^{A,\lambda\rho}\,.
\ee
This term can be written as a total derivative of a non-gauge
invariant vector field $K_{\mu}$, 
composed from the gauge field $G_A^{\mu}$
and, therefore it can be dropped in the context of the perturbation
theory. ${\cal L}_{\theta}$, however, can 
not be neglected in general.
The vector field $K_{\mu}$ is not gauge
invariant, it can have singular behavior at infinity
and those non-trivial topological field configurations of the QCD vacuum 
can lead to physical CP-violating effects. But strong CP-violation
is severely constrained by the data. One such effect would
be the observation of
 electric dipole moment of the neutron.
Using  the experimental upper limit
one gets  
$\theta< 10^{-9}$. It is puzzling why this term is so small.
This question is referred to in  literature as the strong CP-problem
and its resolution leads to the suggestion of the existence of 
 axions~\cite{Peccei:1977hh}.

The classical Lagrangian upon quantization  gets modified:
the perturbative treatment requires that gauge fixing terms,  
 Fadeev-Popov ghost terms are added to the 
Lagrangian and renormalization requires   
 counter terms 
\be
{\cal L}={\cal L}_{\rm classical} + {\cal L}_{\rm gauge-fixing} +
{\cal L}_{\rm ghost} + {\cal L}_{\rm counter\ terms}\,.
\ee
This Lagrangian then uniquely defines the algorithm (Feynman rules)
for calculating  finite physical amplitudes 
in perturbative expansion~\cite{diagrammar}. The quadratic terms
give  the propagators, the trilinear and quartic terms give the
vertices. The form of the counter terms depends on the choice
of   regularization and renormalization  scheme and are obtained
 by calculating
a few ultraviolet divergent self-energy and vertex contribution.
The algorithm is particularly  simple in  the case of 
dimensional regularization, supplemented by the mass independent
$\overline{\rm MS}$ renormalization scheme~\cite{tHooftVeltman:dimreg}.
The Feynman rules and the one loop counter terms can be found 
in many textbooks~\cite{books}.

\subsubsection{Running coupling $\as(\mu)$ and the  $\Lambda$ parameter}
From the explicit
form of the one loop  counter terms one can easily derive
the leading term of the beta function.
It gives the measure of the change of the coupling constant
with the change of the renormalization scale.  
In next-to-leading order of   perturbation theory one obtains
\bea\label{betags}
\mu^2\frac{d\as}{d\mu^2}&=&\beta(\as)=-b_0\as^2-b_1\as^3 + \ldots\\
b_0&=&\frac{11C_A-2n_f}{12\pi}\,, \quad
b_1=\frac{17C_A^2-5C_A n_f-3C_F n_f}{24\pi^2}
\eea
where $\as=g^2_s/(4\pi)$,
 $n_f$ denotes the number of the quark flavors and $C_A$ is
the color charge of the gluons $f_{ACD}f^{BCD}=C_A\delta_{AB}$, 
for $SU(N_C)$ gauge symmetry $C_A=N_C$. 
$b_0$ and $b_1$  are independent from  regularization and renormalization schemes.
Equation~(\ref{betags}) can be easily integrated
\be\label{runningas}
\as(\mu^2)=\frac{1}{b_0\ln{\frac{\mu^2}{\Lambda^2}}}
\left[1-\frac{b_1}{b_0^2}
\frac{\ln \ln \frac{Q^2}{\Lambda^2}}
{\ln\frac{Q^2}{\Lambda^2}}\right] 
\ee
$$
\as(\mu^2)=\frac{1}{b_0\ln{\frac{\mu^2}{\Lambda^2}}}
$$
where $\Lambda$ is an integration constant.
The value of the $\Lambda$ parameter can be extracted
for example from data on 
the scaling violation of deep inelastic scattering (see section 2.1).
Its actual value  $\approx 300 \mev$
gives the measure of the strength of the gluon interaction and the energy
scale at which the coupling constant becomes strong.
By now  a large number of competing methods for extracting
the value of the coupling is available. The results are conveniently normalized
to the scale $\mu=M_Z$ and the world average  is $\as(M_Z)=0.119 \pm 0.003$.
This value is relatively large, 
therefore, the proper scale choice  of the running coupling
 is  far more  important issue in QCD than in QED.
With  appropriate choice of the renormalization scale
one can  avoid the occurrence of 
large logarithms  in the higher order perturbative
corrections.

\subsubsection{Classical versus Quantum Symmetries, Approximate Symmetries}
Local consistent relativistic quantum field theories form a rather limited
class. Full consistency requires renormalizability and gauge symmetry.
One should note in this respect two important features. First, 
with requiring only  
local gauge symmetry, the classical Lagrangian, 
in many cases, possess additional global or discrete symmetries
called accidental symmetries. Secondly,  
 symmetries of the classical theory
can be lost at the  quantum level. We shall consider
two examples. 

If the quarks are  massless, the QCD Lagrangian, is (accidentally) 
scale invariant. 
 The scale invariance of the 
classical theory, however, can not be maintained in the quantum theory.
Quantum fluctuations of the hard modes
are  eliminated by the procedure
of renormalization   defined with  the help of some
scale parameter providing  a hard source of scale symmetry violation.
The coupling is renormalized at a given scale and it  makes  clear 
distinction between the behaviors
at different mass scales. The hidden scale
is introduced
via the running coupling. The 
 $\Lambda$ parameter 
gives  the characteristic scale of the quantum
theory and by definition it is independent from the scale choice
in the running coupling constant. In terms of $\as$ and $\mu$
in leading order 
it is  given as
\be
\Lambda_{\overline{MS}}=\mu e^{-\frac{1}{2b_0\as(\mu)}} \,.
\ee
QCD can be defined non-perturbatively
by formulating it
on a four dimensional Euclidean space-time
lattice~\cite{montvay}. The continuum limit is
obtained in the zero lattice site and large
volume limit. All lattice studies rely on 
the fundamental assumption  that the 
bare coupling constant goes to zero
as given by the perturbative renormalization group 
(see equation~(\ref{runningas})). The validity
of this assumption is by far not trivial but if we accept it then
the $\Lambda_{\rm QCD}$  gives   the
physical scale also for   non-perturbative quantities. In particular,
if we calculate hadron masses, we have to get
\be
m_H=\Lambda_{\rm QCD} c_H
\ee
where $c_H$ is some pure number of order one. 
A quantum theory must have an intrinsic scale. This phenomenon
is referred to in  literature as dimensional transmutation.

The second  example is  the spontaneously broken 
chiral symmetry and the $U(1)_A$ problem. For many purposes
the  quark  masses are negligible if
 $m_q<< \Lambda_{\rm QCD}$.
This is   fulfilled for $m_u$ and $m_d$  and with less accuracy
for $m_s$, while  it is badly broken for $m_c, m_b$ and $ m_t$.
Since color interactions are flavor blind rotating the quark
field in the flavor space will leave the Lagrangian invariant.
With three 
light quarks QCD has accidental global (approximate)
 $U(3)_L\times U(3)_R$  symmetry
\be
{\cal L}_{m_f=0}=
\sum_{f=1}^3\bar{q}^f\hat{D}q^f=
\sum_{f=1}^3
\left(\bar{q}^f_L\hat{D}q^f_L + \bar{q}^f_R\hat{D}q^f_R
\right)\,.
\ee
The  approximate
$SU(3)_L\times SU(3)_R\times U(1)_V\times U(1)_A$ 
symmetry, however, has to be  
broken spontaneously to $SU(3)_V\times U(1)_V$ since
the hadron spectrum  has  only approximate $SU(3)_V$ symmetry (there are
no parity doublets) and
baryon number conservation. 
The pion 
could  successfully be interpreted 
as the Goldstone boson of the corresponding broken
 generators of the axial  $SU(2)_A$ symmetry.  The isospin singlet
pseudoscalar mesons $\eta$ and $\eta^{\prime}$, however,
  are too heavy to be considered
as Goldstone-bosons of the broken $U(1)_A$ symmetry.
This difficulty is referred to  in  literature as the $U(1)_A$ problem.
It has been pointed out, however, by  't Hooft~\cite{tHooft:U(1)}, 
 that the $U(1)_A$ classical symmetry
is lost at the quantum level. The conservation
of the singlet axial current is formally violated
by the Adler-Bell-Jackiw anomaly. 
It was not clear how entire units of axial U(1) charge can
be created from or annihilated into the vacuum. 't Hooft pointed out that
instantons provide the necessary  mechanism: they break the chiral U(1)
symmetry explicitly and $\eta$ and $\eta'$ get 
masses due to instanton contributions.
All this gives  a consistent picture of 
the  strong interactions of the  pseudo-scalar triplet pions
(or with less accuracy of the pseudo-scalar octet mesons).
Using  chiral perturbation theory  quark mass corrections
  can be taken into account~\cite{leutwyler1985} and the spectroscopy
of the pseudoscalar hadrons can be used to extract the
 light quark masses.

\subsubsection{Heavy quark symmetries}
An  approximate symmetry is also obtained  in the infinite
quark mass limit. In the 
 bound states involving heavy quarks,  
 the  heavy quark acts  only 
as a static source of color charge, therefore the physics does not depend
on the flavor of the heavy quark and its spin orientation.
In this limit the corresponding bound states must exhibit an
 $SU(2N_{hq})$ spectrum symmetry where
$N_{hq}$ is the number of the heavy quarks~\cite{neubert}.
 In practice this symmetry  is only useful
 for heavy hadrons containing charm and bottom quarks. 
The top quark decays  too fast  via weak interaction
to form bound states. 
The relations  obtained
in the exact heavy quark limit
can be corrected systematically  by calculating small perturbative       
symmetry breaking
$1/m_Q$ corrections.

\subsection{Basic Electroweak Theory}

\subsubsection{Fermi theory}
The theory of weak interactions started in 1933 with Fermi's 
theory of beta decay. He also suggested  the name
neutrino for the hypothetical particle invented by Pauli in 1930
to explain the continuous energy spectrum of  the electrons (or apparent
non-conservation of energy).

The Fermi theory of weak interactions 
 provides an
exploitation of quantum field  theory outside the realm of
electromagnetism by  describing processes when 
electrons, neutrinos and atomic nuclei are created and annihilated.
Fermi's original interaction involves two vector currents in analogy
with the electromagnetic 
interaction describing electron-electron scattering.
 The correct form
of the interaction, however, became clear only after the discovery
of parity violation in 1957~\cite{wu} and its theoretical interpretation
by Lee and Yang~\cite{leeyang} which lead to the proposal that 
the Lagrangian of weak   interactions  is
 given by the products of 
(V-A)  currents
\be
{\cal L}_F(x)=\frac{G_F}{\sqrt{2}}
  \bar{p}(x)\gamma^{\alpha}(g_V-g_A\gamma_5)n(x)
\bar{e}(x)\gamma_{\alpha}(1-\gamma_5)\nu(x)+ {\rm h.~c.}
\ee
where the vector coupling of the nucleon
is slightly smaller than one and is given
by the Cabibbo angle $g_V=\cos\theta_C\approx0.97$.
 The  ratio of the axial
to vector couplings of the nucleon is known  from the
study of beta-decay with 
total angular momentum transitions $\Delta J=0,1$\, giving 
$g_A/g_V=-1.2573\pm 0.0028$~\cite{PDGcaso}~. 
This Lagrangian can be used to calculate the neutron lifetime in
 leading order of perturbation theory 
in terms of the Fermi coupling $G_F$.  From the  experimental value of the 
neutron lifetime
$\tau=887.0\pm 2.0$\,sec one obtains a first estimate of the
 value of the Fermi constant
 $G_F\approx (250\gev)^{-2}=1.6\times 10^{-5}$
 $\gev^{-2}$. The theory is not renormalizable and 
the interaction is weak at low energies.

With the discoveries of the pion, the muon and strange hadrons
 the V-A structure of weak interactions has been established in a variety
 of experiments. Further progress has been made with the discovery 
 that the 
electron and muon number are separately conserved and that  
the  neutrinos  associated with the muons are new particles.
The data have indicated  that
the strength and form of the four fermion interactions  between
fermionic doublets $(\Pp,\Pn)$, $(\Pe,\Pne)$, $(\Pmu,\Pnmu)$
is universal
 in particular the muon decay is described by the Lagrangian as
\be
{\cal L}_{\mu}(x)=\frac{G_F}{\sqrt{2}}
  \overline{\nu}_{\mu}(x)\gamma^{\alpha}(1-\gamma_5)\mu(x)
\overline{e}(x)\gamma_{\alpha}(1-\gamma_5)\nu_e(x)+ {\rm h.~c.}
\ee
This  interaction allows for   an important
non-renormalization theorem: the photonic corrections to this
transition are finite in all orders of perturbation theory
\cite{bermansirlin}. The leading corrections have been
calculated 20 years ago~\cite{kinoshita}\,,      
the ${\cal O}(\aopi)^2$ term
has  been obtained 
by van Ritbergen and Stuart only very recently \cite{vanRitbergen:1999fi}.
The muon lifetime is then given by the theoretical expression
\begin{equation}\label{GF}
\frac{1}{\tau_{\mu}} =\frac{G^2_F m_{\mu}^5}{192\pi^3}\left(1-\frac{8m_e^2}
{m^2_{\mu}}\right)\\ \nonumber
\left[1+1.810\left(\frac{\alpha}{\pi}\right) + 
(6.701\pm 0.002)\left(\frac{\alpha}{\pi}\right)^2+ ...\right]\,.
\end{equation}
This equation offers a convenient  definition
of the Fermi-coupling $G_F$  by assuming that 
 the non-photonic corrections are 
all  lumped into $G_F$ in a way that 
it can be considered  a  physical quantity.
Using the measured value of $\tau_{\mu}$ $\cite{PDGcaso}$ we get
\begin{equation}
G_F=(1.16637\pm 0.00001)\times 10^{-5}\gev^{-2}\,.
\end{equation}

\subsubsection{Weak isospin and hypercharge}
In seeking  analogy  between electromagnetism and week
interaction, the four-fermion interactions can
be considered as the effective low energy theory of a charged massive vector
boson interacting with the charged chiral current
\bea
{\cal L}_I&=&-
\frac{g}{2\sqrt{2}}
W^-_{\alpha}J^{+\,\alpha} + {\rm h.c.}\\ \nonumber
J^+_{\alpha}&=&\left[\overline{\nu}_{e}(x)\gamma_\alpha
(1-\gamma_5)e(x) + \ldots\right]\\ \nonumber
G_F&=&\frac{g^2}{8\MW^2},\quad  \MW\le\ 110\gev,\ \ \  {\rm if}\ \ g<1
\eea
where $W^{\pm}=W^1\pm iW^2$. It is natural to
consider the charged current  as the 
 charged component of the weak isospin $SU(2)_L$ current
\bea
\frac{1}{2} J^i_{\alpha}&=&\overline{N}_{L}(x)\gamma_{\alpha} T^i
 N_L(x)+\overline{L}_{L}(x)\gamma_{\alpha} T^i L_L(x) \nonumber
\\ 
\hspace{-1cm}
 N_L(x)&=&
{\cal P}_L \left( \begin{array}{c} p(x)\\ n(x) \end{array}
\right) \,, 
\quad  L_L(x)={\cal P}_L
 \left( \begin{array}{c} \nu(x)\\ e(x) \end{array}
\right)\quad {\cal P}_L=\frac{1}{2}(1-\gamma_5)
\eea
where $T^i=\tau^i/2$ is the $SU(2)$ generator in the fundamental
representation.
This assumption, however, implies necessarily that
in addition to  electromagnetism 
 weak neutral current must exist since
$$
[T^+,T^-]=2 T^3 \ne Q\,.
$$
where $T^{\pm}= T^1\pm i T^2$.
 Actually, one assumes that the $SU(2)_L$ doublets and singlets 
carry a diagonal hypercharge quantum numbers such that
\be
Q=T^3+Y
\ee
is fulfilled. Furthermore,  
since hadrons are composite state of quarks, 
 the weak hadronic currents
have to be given not in terms of nuclei but quark doublets.
The $SU(2)_L\times U(1)_Y$ quantum numbers of left and right handed
 quarks and leptons
are listed in Table 1. 
The spin half matter field form three
identical quark-lepton families. It is convenient to
classify the  matter fields in terms of left handed Weyl spinors.
This is possible since 
under CP conjugation a right-handed spin half fermion
is transformed into 
 a left-handed antifermion. 
We can group
the fundamental spin half particles  into
 a reducible multiplet of 
doublet fermions   and  singlet antifermions. 
One  quark-lepton family is composed from 15 left-handed
Weyl fermions
grouped in 5 irreducible components
\be\label{reducefermions}
\psi^f_L=\left[ Q^f_L(3,2,1/6),U^f_{cL}(3,1,-2/3), D^f_{cL}(3,1,1/3),L^f_L(1,2,-1/2),
E^f_{cL}(1,1,1)\right],f=1,2,3
\ee
where the first two numbers in the ordinary parenthesis 
 are the dimensions of the SU(3)
and  SU(2) representations, respectively 
the third number is the  value of the hypercharge
and $f$ is the family label.
The corresponding Dirac spinors  will be labeled
as 
$
\psi_{f_{\chi}}
$
 where $\chi$ runs over the values $\chi=U,D,E,N$ 
and
\be\label{labels}
f_U=u,c,t\,, \ \    f_D=d,s,b\,, \ \  f_E=e,\mu,\tau\ \  {\rm and}\ \ 
 f_N=\Pne,\Pnmu,\Pntau 
\ee
where $f_{\chi}$  is again the family label but for a given  component
of the families.
\begin{table}
\begin{center}
\vskip0.2cm
\begin{tabular}{ |c c|c|c|c|c|c|c|c|c|} \cline{1-10} 
 & \multicolumn{3}{c|}{families} & \multicolumn{1}{c|}{color} 
 & \multicolumn{1}{c|}{$T^3_L$} & \multicolumn{1}{c|}{$Y_L$}
 & \multicolumn{1}{c|}{$T^3_R$} & \multicolumn{1}{c|}{$Y_R$}
 & \multicolumn{1}{c|}{$Q$} 
 \\ \cline{1-10} 
  & $\left(\begin{array}{c}u\\ d\end{array}\right)$
  & $\left(\begin{array}{c}c\\ s\end{array}\right)$
  & $\left(\begin{array}{c}t\\ b\end{array}\right)$
  & $\begin{array}{c}3\\ 3\end{array}$
  & $\begin{array}{c}1/2\\ -1/2\end{array}$
  & $\begin{array}{c}1/6\\ 1/6\end{array}$
  & $\begin{array}{c} 0\\ 0\end{array}$
  & $\begin{array}{c}2/3\\ -1/3\end{array}$
  & $\begin{array}{c}2/3\\ -1/3\end{array}$
 \\ \hline 
  & $\left(\begin{array}{c}\Pne\\ \Pem\end{array}\right)$
  & $\left(\begin{array}{c}\Pnmu\\ \Pmum\end{array}\right)$
  & $\left(\begin{array}{c}\Pntau\\ \Ptaum\end{array}\right)$
  & $\begin{array}{c}1\\ 1\end{array}$
  & $\begin{array}{c}1/2\\ -1/2\end{array}$
  & $\begin{array}{c}-1/2\\ -1/2\end{array}$
  & $\begin{array}{c} 0\\ 0\end{array}$
  & $\begin{array}{c}0\\ -1\end{array}$
  & $\begin{array}{c}0\\ -1\end{array}$
 \\  \hline
\end{tabular}
\end{center}
\caption[dummy]{\small Quantum numbers of the fundamental fermions 
 \label{tab:QNfunferm}}
\end{table}

\subsubsection{Towards Yang-Mills theories}
The universality of the interactions, the weak isospin structure and 
the analogy with QED pointed  to  the Yang-Mills    theory
 with gauge group of
 $SU(2)_L\times U(1)_Y$~\cite{glashowSM, weinbergSM}.
The symmetric part of the Lagrangian density is given
in terms of the two gauge coupling constant $g$ and $g^{'}$
\be\label{YMLagsym}
{\cal L}_{\rm ew}= -\frac{1}{4}W^{i,\mu\nu}W^i_{\mu\nu}
 -\frac{1}{4}B^{\mu\nu}B_{\mu\nu} 
+ 2 \sum_{f=1}^3\overline{\psi}^f_L\gamma_{\mu}D^{\mu}\psi^f_L \,.
\ee
where $D_{\mu}$ is 
the covariant derivative 
\be\label{covderiv}
D^{\mu}=\partial^{\mu}+ig~ t^i~ W^{i,\mu}+ig^{\prime}~Y~B^{\mu}
\ee
All terms containing the gluon fields are dropped and
$t^i$ is the  $SU(2)_L$ matrix of the reducible fermionic
representation $\psi_L^f$. The photon field is the linear
combination of $W_3$ and $B$ coupled to the electromagnetic
current 
\be
A_{\mu}=-\sin\theta_W W^3_{\mu} + \cos \theta_W B_{\mu}\\  
\ee
with
\be
\tan\theta_W=\frac{g^{\prime}}{g} 
\,, \quad e=g\sin\theta_W\,.
\ee
The Z-boson field is the  orthogonal combination
\be
Z_{\mu}=\cos\theta_W W^3_{\mu} + \sin \theta_W B_{\mu}\\  
\ee
 coupled to the weak neutral current.
The interaction terms of the fermions are 
\bea
{\cal L}_{I_f}&=&-\biggl(
\frac{g}{2\sqrt{2}}J^+_{\mu}W^{-,{\mu}}
+\frac{g}{2\sqrt{2}}J^-_{\mu}W^{+,{\mu}}\nonumber 
+\frac{g}{2\cos\theta_W}
J^{\rm NC}_{\mu}Z_{\mu}
+e J^{\rm elm}_{\mu}A^{\mu}\biggr)
\eea
with currents defined in terms of Dirac spinors
\bea
J^+_{\mu}&=&\sum_{f_U,f_D}
\overline{\psi}_{f_U}\gamma_{\mu}(1-\gamma_5)V^{f_Uf_D}_{CKM}
\psi_{f_D}
+\sum_{f_E}
\overline{\psi}_{f_N}\gamma_{\mu}(1-\gamma_5)\psi_{f_E}
\nonumber \\
J^{\rm NC}_{\mu}&=&\sum_{f_{\chi}}
\overline{\psi}_{f_{\chi}}
\gamma_{\mu}(v_{\chi}-a_{\chi}\gamma_5)\psi_{f_{\chi}}
\nonumber \\
J^{\rm em}_{\mu}&=&\sum_{f_{\chi}}
\overline{\psi}_{f_{\chi}}\gamma_{\mu}Q_{\chi}\psi_{f_{\chi}}
\eea
where $f_{\chi}$  are the labels defined in equation~(\ref{labels}),
the color labels and spinor labels are suppressed and 
$V_{CKM}^{ff{'}}$ is the CKM-matrix (see subsection~\ref{CKM}).
The requirement of non-Abelian gauge symmetry leads to 
 the universality of the
gauge boson interactions and  predicts  the neutral
current couplings
\be\label{afvf}
v_{\chi}=T^3_{\chi,L}-2Q_{\chi}\sin^2\theta_W\,, \quad a=T^3_{\chi,L}
\quad \chi=U,D,E,N\,.
\ee
The chiral gauge symmetry  of the 
 Lagrangian~(\ref{YMLagsym}) forbids mass terms both
for the gauge bosons and  the fermions. 
Adding mass terms by hand is disastrous since it destroys gauge invariance.
Because of this in the early sixties these theories have not
been taken seriously and the  successful
predictions for the neutral currents
were considered to be very vague.
The fact that the low energy effective
theory  was rather successful in explaining the
charged current data   implied
small  correction terms  and gave an experimental hint 
that somehow  massive renormalizable 
Yang-Mills theories must exist~\cite{Veltmanhint}.
The breakthrough came with the  solid theoretical
 understanding of 
the renormalizability  of 
the Yang-Mills theories~\cite{thooft1,thooft2} and the mechanism of 
mass generation.

\subsubsection{Higgs mechanism \label{higgsmechanism}}
%
The difficulty with the mass terms and its resolution
can be understood already in the case of abelian theories.
Massless spin one particles have only two spin degrees
of freedom, the longitudinal component does not
contribute to the kinetic energy and the free theory
is gauge invariant. Keeping the interactive theory
gauge invariant, the longitudinal components remain decoupled and
one gets renormalizable theories. 
Adding even an infinitesimal mass term is disastrous:
the longitudinal component of the gauge bosons becomes physical and
 it  destroys unitarity.
The trouble is related to the number of degrees of freedom:
the massive gauge bosons have three spin states, therefore, 
 the massless theory can not be obtained simply as the massless limit
of a massive theory.
At high
energies, however,
 the longitudinal component behaves like a scalar particle suggesting
that perhaps 
the gauge symmetry may be maintained if we add 
scalar particles to the theory. The gauge transformation
rules then will also involve the scalar field.
This was the crucial  observation 
of Higgs~\cite{Higgs:1964pj}
 Brout and Englert~\cite{Englert:1964et}
 leading  to the discovery
of the Higgs mechanism. 
Even if  the
energetically preferred value of the scalar field
is not equal to zero, 
 the Ward-Takahashi identities required
by local gauge invariance can be maintained.
If  the ground state $< \phi > $ is non-vanishing, 
without the requirement of local symmetry, 
 we get spontaneous symmetry breaking with
massless Goldstone bosons associated with each broken generators.
In gauge theory  
at high energies when  masses are negligible we have
massless Goldstone bosons and massless gauge bosons.
At low energies, however,   
 the Goldstone bosons disappear from the theory: they provide
the longitudinal component of the massive gauge bosons since the number
of degrees of freedom of the theory has to be preserved.
This feature of gauge theories coupled to Goldstone bosons  is called
the Higgs mechanism. One can obtain massive gauge 
 bosons  by supplementing  the Lagrangian with some new sector 
providing us with the appropriate  
 Goldstone bosons.

The Standard Model is defined with 
the simplest realization of the Higgs mechanism~\cite{weinbergSM}:
 one adds to the theory
one  scalar doublet 
with appropriate hypercharge $Y(\Phi)=1/2$
\bea
 \Phi=\left( \begin{array}{c}
\phi^+ \\ \phi^0 \end{array} \right) 
\eea
with gauge kinetic energy term and self-interactions
\be
{\cal L}_{\Phi}(x)=\left(D_{\mu}\Phi\right)^{\dag}D^{\mu}\Phi + 
\mu^2\Phi^*\Phi-\lambda\left(\Phi^{\dag}\Phi\right)^2
\ee
and Yukawa couplings
\bea
{\cal L}_{\rm Yukawa}(x)&=&
\sum_{ff^{'}}\lambda^U_{ff^{'}}\left(\overline{Q}_{Lf}
\tilde{\Phi}\right)u_{Rf^{'}}+
\lambda^D_{ff^{'}}\left(\overline{Q}_{Lf}
\Phi\right)d_{Rf^{'}}\nonumber\\
&&\hspace*{-0.3cm} + \lambda^E_{ff^{'}}\left(\overline{L}_{Lf}
{\Phi}\right)e_{Rf^{'}} + {\rm h.~ c. }
\eea
where $Q_{Lf}$ and $L_{Lf}$ denote the quark and lepton doublet
Weyl spinors for family $f$ and 
$\lambda^u_{ff^{'}}$, $\lambda^d_{ff^{'}}$, $\lambda^e_{ff^{'}}$
denote complex coupling matrices in the family space.
There is  no Yukawa coupling for neutrinos, since it is assumed
that in nature 
only left-handed neutrinos exist.
Assuming $\mu^2 > 0$ there is a circle of degenerate minima at
\be
\abs{\Phi}^2=\frac{\mu^2}{2\lambda}\equiv \frac{v^2}{{2}}\,.
\ee
The excitations along the circle correspond to the Goldstone bosons.
The local gauge transformations, however, also rotate $\abs{\Phi}$ 
 along the circle.
One can choose gauge condition in a way that the scalar field points
(at least in leading order) to a fixed direction (unitary gauge).
 If we had 
dealt with global symmetry we would have gotten spontaneous symmetry
breaking as the vacuum is not symmetric 
with non-vanishing scalar field. The vacuum, however, does not break the local
gauge invariance. Any state in the Hilbert space that  is
not invariant under local gauge transformations is unphysical.
With choosing unitary gauge  
\bea\label{phishift}
 \Phi(x)=\frac{1}{\sqrt{2}}\left( \begin{array}{c}
0 \\ h(x)+v \end{array} \right) 
\eea
 local gauge
invariance is not broken 
but one rotates away the three unphysical components of the scalar
doublet field $\Phi(x)$. The $h(x)$  field describes 
the neutral Higgs boson remaining in the physical spectrum.  
Rewriting the
Lagrangian in terms of $h(x)$ and $v$ we obtain  mass terms  for
the gauge bosons 
\be\label{gbmassmatrix}
{\cal L}^{VB}_m=\abs{D_{\mu}<\Phi>}^2=\frac{v^2}{4}\Biggl [
g^2 W^+_{\mu}W^{-\mu}
+\frac{1}{2}(W_{\mu}^3,B_{\mu})
\left( \begin{array}{cc}
g^2 &g g^{'}\\ g g^{'}& g^{'2} \end{array}\right)
\left( \begin{array}{c} W^{3\mu}\\B^{\mu}\end{array}\right)
\Biggr]\,.
\ee
In terms of the $A_{\mu}$ and $Z_{\mu}$ fields
the mass matrix of the neutral gauge bosons becomes
diagonal and one gets
\be
 \MW=\frac{1}{2}gv\,,\quad \MZ=\frac{1}{2}\sqrt{g^2+g^{'2}}\,,
\quad m_{\gamma}=0\,.
\ee
For the mass of the Higgs boson we obtain
\be
\MH^2=2\lambda v^2
\ee
therefore 
the strength of the 
self-interaction of the Higgs boson can be expressed in terms
of the Higgs and gauge boson masses and the gauge coupling 
\be\label{selfcp}
\lambda=\frac{\MH^2}{8\MW^2}g^2\,.
\ee
The gauge symmetry  uniquely defines the 
coupling of the gauge bosons  
to the Higgs boson allowing to predict for example 
 the value of the half-width of the Higgs boson
\be
\Gamma(h\to\PWp\PWm)=\frac{g^2\MH^3}{64\pi\MW^2}
\sqrt{1-4x_h}(1-4x_h+12x_h^2)\,, \quad x_h=\frac{\MW^2}{\MH^2}\,.
\ee
 With increasing $\MH$ it grows  as  $\MH^3$. In particular
for  $M_H\approx 1\tev$ we get  $\Gamma(h)\approx \MH$ indicating the  
difficulty with the validity of perturbative unitarity in  case of
a heavy Higgs boson.

With substituting the shifted field~(\ref{phishift}) in
the Yukawa coupling of $\Phi$ to fermions we get the mass matrices
of the fermions and their couplings to the Higgs boson.
The physical fermion states are obtained  by
 diagonalizing the mass matrices (with biunitary rotations)
\be
\frac{v}{\sqrt{2}}
U({\chi})^{\dag}_L
\lambda^{\chi}U({\chi})_R\,,
=
{\cal M}_{\rm diag}^{\chi}\,,\quad   \chi=U,D,E\,.
\ee
$\chi=N$ does not occur since it is assumed that 
right handed neturinos do not exist.
The diagonal element of 
 ${\cal M}_{\rm diag,ff{'}}^{\chi}=m^{\chi}\delta_{ff^{'}}$
gives the mass values and $f$ runs over the three families.
This diagonalization produces    three 
important physical results. 

 First,   the couplings of the Higgs boson to fermions
  are flavor diagonal and proportional to the fermion mass
\be
{\cal L}^Y_h(x)=- \sum_{\chi,f}\frac{g m_{f_{\chi}}}{2\MW}\overline{
\psi}_{f_{\chi}}(x){\psi}_{f_{\chi}}(x)\, h(x)\,,
\ee
 therefore, the coupling
of the Higgs bosons to light fermions is very weak. This makes
its experimental search extremely difficult.

Secondly, the  charged  current of quarks is not flavor diagonal 
\be
J^-_{\mu}(x)=
\sum_{f_U,f_D}
\overline{\psi}{f_U}(x)
V^{f_Uf_D}_{\rm CKM}
\gamma_{\mu}(1-\gamma_5)\psi_{f_D}(x)
\,, \quad f_U=u,c,t\,, \quad f_D=d,s,b
\ee
where
 $V_{\rm CKM}=U^{\dag}(U)_LU(D)_L$ denotes
the Cabibbo-Kobayashi-Maskawa matrix.
 In the  charged current of
quarks six fields  with 5 physically irrelevant
independent phases are involved 
(with one relevant phase for $U(1)_Y$), 
therefore four phases of  unitary CKM matrix  
 can be rotated away and
we end up with $3^2-5=4$ physically relevant parameters.
If the neutrinos are massless, one can arbitrarily
rotate the neutrinos so that the charged lepton
current remains diagonal. Recently, experimental evidence
has been obtained for neutrino oscillations and so
for neutrino masses~\cite{kamiokande}. 
The pattern for massive  neutrinos is  more
complicated than those for massive quarks as the electrically neutral
 neutrinos can   have  Dirac and/or Majorana mass terms.
In the minimal Standard Model
with one Higgs doublet, however, only the Dirac mass term is possible.

Thirdly,  after
 the Higgs mechanism the neutral current remains obviously
flavor diagonal (GIM mechanism~\cite{GIM}), therefore in leading
order  there are no flavor changing neutral current transitions.
Before the discovery of the charm quark this feature was  less obvious.
In higher order, as a result of virtual flavor changing charged current
exchanges, flavor changing neutral current transitions are allowed
but suppressed strongly by the smallness of  the Fermi coupling.

\subsubsection{ CKM matrix \label{CKM}}
The elements of the CKM matrix are denoted conveniently
as $V^{ff{'}}_{\rm CKM}$ with $f=u,c,t$ and $f^{'}=d,s,b$. 
As we noted above, they can be described in terms of four independent
parameters. The values of these 
have to  be extracted from the data.
Although the  independent parameters of the CKM matrix 
are fundamental
parameters of the theory, there is  considerable freedom in their
definite choice. According to the data
 the matrix elements   are large in the diagonal and
they get smaller and smaller as we move away from it.
Wolfenstein~\cite{wolfensteinpar}
 has suggested a convenient parametrization which
takes this hierarchical behavior into account
\bea
V^{\rm CKM}=\left(
\begin{array}{ccc}
 1-\lambda^2/2            & \lambda           & A\lambda^3(\rho-i\eta)\\
 -\lambda                 & 1-\lambda^2/2    & A\lambda^2 \\
 A\lambda^3(1-\rho-i\eta) &-A\lambda^2         &1
\end{array}
\right) + {\cal O}(\lambda^4) \,.
\eea
Experimentally $\lambda\approx 0.22$, $A\approx 0.82$
$\sqrt{\rho^2 + \eta^2}\approx 0.4$ and $\eta\approx 0.3$~\cite{PDGcaso}.
Since $\eta$ is non-vanishing  CP-violation  occurs both in the 
neutral kaon mass matrix and in the  transition amplitudes,
therefore, the CKM model of  CP-violation
 is milliweak. Recent experiments also confirmed the presence
of CP-violation  in the decay amplitudes of kaons into pions
in agreement  with the prediction of the Standard Model.
Unfortunately, the theoretical predictions have large
theoretical error due to non-perturbative QCD effects~\cite{buras},
 therefore 
the agreement gives only a qualitative confirmation.
The CKM predictions  of CP-violating effects in terms 
of a single CP-violating parameter depend sensitively
on  the assumptions  that the Higgs sector is minimal 
and that we have only three fermion families.
 If the Higgs sector
is more complicated or there are  additional
heavy fermions the predictions of the Standard Model
for CP violation effects will fail. 
This is why the precision  test of the CKM predictions for
  CP-violation is so important.
The CKM matrix is unitary in the Standard Model but
 is not necessarily unitary in its extensions. 
 It is of great interest to test the  relations
among the CKM matrix elements required by unitarity.
A particularly interesting relation is the prediction
\be\label{triangle}
V_{td}V^*_{ud}+V_{ts}V^*_{us}+V_{tb}V^*_{ub}=0\,.
\ee
All the three terms at the left-hand side 
 are proportional to $A\lambda^3$.
 Up to  this overall factor
the three terms are equal to $1-\rho + i\eta$,\   $\rho-i\eta$,\  $1$
 and they form a triangle 
(Bjorken) in the complex plane.  The  surface of the triangle is equal
to the Jarlskog invariant~\cite{cecilia}
\be
J\approx A\lambda^3\eta\,.
\ee
If the CKM matrix conserves CP,
the unitarity triangle shrinks  to  a line.
The observation of  CP-violation in b-decays in the near future 
 will provide  a  decisive
 test on the validity of 
the CKM model of CP-violation and    the fermion-mass generation
mechanism of the Standard Model.

\subsubsection{Custodial symmetry}
We have already noted in section 1.18 that in the case of a simple
matter field content  the requirement of local gauge symmetry and
 renormalizability may lead to a
 Lagrangian density with accidental global symmetries.
The Higgs sector of the Standard Model
 has an  accidentally   global
$SU(2)_L\times SU(2)_R \equiv SO(4)$ symmetry. 
The non-vanishing vacuum expectation value of the Higgs field
breaks it
down spontaneously to the diagonal $SU(2)_V$ global symmetry (called
custodial symmetry).
In the limit  $g^{'}=0$ the gauge interaction preserves 
this  symmetry  and   in this limit the massive
gauge bosons must form a degenerate triplet representation
of $SU(2)_V$. 
Non-vanishing
$g^{'}$ coupling leads to the mass splitting
$\MW=\cos\theta_W\MZ$ (see equation~(\ref{gbmassmatrix})).
This relation therefore remains valid for any Higgs mechanism
which respects custodial symmetry. The
 Yukawa couplings also violate the custodial
symmetry if the mass values of the up and down  components of a fermion
doublets 
are not degenerate. Since the top-bottom mass splitting
is large,  virtual  top and bottom quark contributions
give large corrections to the leading
order  value of the $\rho$ parameter
$
\rho=\MW/(\cos\theta_W\MZ)\,.
$

\subsubsection{Cancellation of chiral gauge anomalies}
Classical chiral gauge symmetries may be broken in the quantum
theory by triangle anomalies.
Chiral fermions are massless but renormalization requires the
regularization of the  theory which 
 necessarily introduces masses for the fermions. It may happen that the 
chiral symmetry of the classical theory will not survive in the
quantum theory.
This is disastrous for the chiral gauge theories because 
 the gauge symmetry is  broken and the theory makes
no sense. Fortunately, all  terms which break chiral
symmetry have very simple origins as 
they  come from simple fermionic triangle diagrams 
coupled to  vector and  axial vector currents.
 Therefore,  the anomaly is
 proportional to the difference of  the trace of coupling
matrices 
of the left-handed 
 and right-handed fermions (masses are negligible
in the ultraviolet limit). A chiral gauge theory is only 
meaningful if the chiral anomalies cancel each other.
One can easily see that the condition of anomaly cancellation
in the Standard Model is that the
 sum of the charges of the fermions  vanishes.
In the Standard Model this is fulfilled individually 
 for each family  
\be\label{chargetrace}
Tr{Q}=3(Q_u+Q_d)+Q_e=0\,.
\ee
This simple result follows  because  the group $SU(2)$ is anomaly free.
The condition~(\ref{chargetrace}) forms 
 an important bridge between the electroweak
sector and the strong sector: without  quarks the lepton
sector is anomalous and the quarks must come 
 in three colors.
We note the problem of charge quantization.
It is a phenomenological fact that
the charges of the proton and the positron
are equal to each other within very 
high experimental accuracy. The relation
$Q=T_3+Y$, however,  would allow arbitrary 
relation since the $U(1)_Y$ charge is not
quantized. By choosing
the value of $Y$ consistently with value of 
the charges of the positron  and the proton
 we get the  
anomalies cancelled. This indicates that   charge quantization
and anomaly cancellation may be connected. 

In the case of global
symmetries the anomalies do not destroy renormalizability, 
but the quantum theory will not be symmetric.
For example the strong chiral-isospin symmetry
forbids the decay  $\pi^0\to \gamma\gamma$, but in the quantum
theory this symmetry is violated by the triangle anomaly and the
decay is allowed. The anomaly is proportional to $Tr(Q^2T_3)=
N_c(Q_u^2-Q_d^2)$. This result  played a crucial role in the  discovery
 of color.

\subsubsection{Accidental continuous global symmetries}
It is a  success  of the
Standard Model that 
 the requirement
of local gauge invariance and renormalizability
leads to  accidental global symmetries.
 The origin of these
symmetries is  the large 
$U(45)$ symmetry  of the Lagrangian of the 45 Weyl fermions
 of the Standard Model.
 The gauge interaction breaks
down this symmetry to three copies of 5 irreducible
$SU(3)\times SU(2) \times U(1)$ representations 
(see equation~(\ref{reducefermions})) 
which still has  $U(3)^5\times U(1)$ global symmetry.
This is broken by the Yukawa coupling to $U(1)^4$. Because
of the CKM matrix in the quark sector,  only one 
phase rotation survives:
 each quark is rotated with the same phase and
each antiquark field is rotated with the opposite phase.
Since we do not have a CKM matrix in the lepton sector
we can have individual phase rotations of leptons for each family.
These symmetries lead to the $\it exact$  baryon number $B$ and to individual
lepton number conservations for each family 
$ L_e$, $L_{\mu}$ and $L_{\tau}$.
According to the data 
these conservation laws are valid to very high precision~\cite{PDGcaso}.
I have to note that
we do not consider these symmetries to be absolute.
They  are violated even within the Standard
Model by instanton contributions~\cite{tHooft:instantons}. 
The baryon current   coupled to  two gauge currents
via the anomalous triangle diagrams 
is necessarily  anomalous. The condition~(\ref{chargetrace})
requires that $B-L$ is conserved. The instantons can absorb
baryon and lepton numbers.
Also, 
considering the Standard Model as an effective low energy
field theory  with all possible
 higher dimensional non-renormalizable gauge invariant
operators baryon and lepton number can be violated.
The baryon and lepton number violation is strongly suppressed
by power corrections.
The search for such effects
 is an important tool
for testing  the range of validity of the Standard
Model.

\section{ PRECISION CALCULATIONS}
\newcommand\epb{\overline{\epsilon}}
\subsection{Testing  QCD}

QCD  is  asymptotically free, therefore  the
physical phenomena
at short distances and at finite time intervals
 may  be in principle  subject to perturbative treatment
in terms of  weakly interacting quarks and  gluons.
It is crucially important for collider physics that  the 
 perturbative description  is valid
  for large momentum transfer reactions
since
 perturbation theory is the only systematic method for calculating
scattering cross sections directly from the QCD Lagrangians.
The application of perturbative QCD to scattering phenomena with large
momentum transfer, however,  is not  straightforward. It is not
obvious a priori that
the short and long distance properties  can be   meaningfully separated.

In general the cross sections of 
scattering processes in perturbative QCD with
massless quarks and gluons  are  singular
due to the presence of soft and collinear contributions.
Fortunately, in simple cases when only one hard scale is involved
like the total cross section of $\epem$ annihilation to hadrons, 
 deep inelastic scattering and the   Drell-Yan
processes, one can prove in all order of the perturbation theory
that  all the infrared sensitive contributions given
by soft and collinear parton configurations are cancelled 
except some remaining collinear singularities~\cite{kinoshitalee,
amativen13,bodwin14,colsop11}. 
They are, however,   universal
and can be factored into the parton distribution functions
of the incoming  hadrons or into the  fragmentation functions of final hadrons.
The fundamental assumption of the QCD improved parton model
is  that 
this theorem  remains valid after including
 non-perturbative effects   up to
some power corrections of  ${\cal O}(\Lambda_{QCD}/Q)$ where 
$Q$ denotes the hard scale of the process.
Furthermore, it is assumed that the theorem 
remains valid for any infrared safe quantities.
A physical observable is called infrared safe if
its value calculated in the perturbation theory is not sensitive
to the emission of additional soft gluons or the splitting
of a hard parton into two collinear partons.

In the QCD improved parton model 
the incoming hadrons
are considered as wide band  beams of hadrons  with well defined
momentum distributions. Infrared safe 
 hard scattering cross sections 
of hadrons are  calculated in the  perturbation theory in terms of partons.

\subsubsection{Hadron production in $\epem$ annihilation} 

In simple inclusive reactions, 
such as  the total cross section of $\epem$ annihilation into
quarks and gluons,
 the soft and collinear contributions
cancel~\cite{kinoshitalee} (KLN theorem).
Therefore, the cross section is free from infrared
singularities and can be calculated in power series
of the effective coupling 
\be
R = \frac {\sigma(e^+e^- \to {\rm hadrons})}{\sigma (e^+e^- \to \mu^+
\mu^-)} = \overbrace{(1 + \frac {\alpha_s}{\pi}+...)}^{\bar R} 3\sum_2
e_{q}^{2}
\ee
where 
\be\label{Rhad}
\bar R = 1 + \frac{\alpha_s(\mu)}{\pi} + \biggl(\frac
{\alpha_s(\mu)}{\pi} \biggr)^2 \bigl[ \pi b_0 \ln \frac{\mu^2}{s} + B_2
\bigr] + ...
\ee
$\as$ is the running coupling constant, 
 $\mu$ is the renormalization scale, $B_2$ is a known constant
given by the NNLO calculation~\cite{nnloepem} 
 and $b_0$ is the first coefficient
in the beta function given in equation~(\ref{betags}).
The explicit $\mu$ dependence in (\ref{Rhad})
 is cancelled by the $\mu$ dependence
of the running coupling constant to ${\cal O}(\as^3)$.  In general, 
the  truncated series is $\mu$-dependent but the $\mu$ dependence is 
  order of
${\cal O}(\as^{(n+1)})$ if the cross section is calculated to 
${\cal O}( \as^{n})$.
 
The result obtained for partons can be applied to hadrons assuming that 
the unitarity sum over all
hadronic final state can be replaced with the unitarity sum
over all final state quark and gluons 
\be
\sum_h | h > < h | = \sum_{q,g}
 |{\rm gluons, quarks}>
 <{\rm gluons, quarks}|  \ .
\label{fundassump}
\ee
This assumption 
is only  valid 
if the annihilation energy is much larger than the quark
and hadron mass values
and if the annihilation energy  $Q$ is 
in a region far from resonances and thresholds (or we smear over
the threshold and resonance regions).

 The KLN theorem remains
valid also for integrating over final states in a limited phase space
region, as in the case of jet production.
The Sterman-Weinberg two-jet cross section~\cite{stermanweinberg}
 is defined by requiring
that all the final state partons are within a back-to-back cone
 of size $\delta$, 
provided their energy is less than $\epsilon \sqrt{s}$.
At NLO
\bea
\sigma_{\rm 2jet} &=& \sigma_{\rm SW}
 (s,\varepsilon,\delta)
 \nonumber \\
&  = &\sigma_{\rm tot}
 - \sigma_{q \bar qg}^{(1)} ({\rm all}\  E > \varepsilon \sqrt{s}, 
{\rm all}\ \theta_{ij} > \delta)
\nonumber \\ &=&
\sigma_0 
\biggl[ 1 - \frac{4\alpha s}{3\pi}
    \bigl(4 \ln 2 \varepsilon
\ln \delta + 3 \ln \delta - 5/2 + \pi^2/3\bigr)
\biggr]
\eea
where $\sigma_0=4\pi\alpha^2/3s$.
We can easily see that the jet
definition is infrared safe and therefore 
the cancellation theorem 
remains valid. The dependence of the cross section on the jet defining
parameters $\epsilon,\delta$ is physical since  the same parameters 
have  to be used 
 in the measurements of jet cross sections  
when  jets are  defined in terms of hadrons.

\subsubsection{Hard scattering with hadrons in the initial state}
In infrared safe
quantities for processes with partons in the initial state,
the initial state collinear singularities are not cancelled but they
are universal ( process independent) in all 
orders in the perturbation theory~\cite{colsoprevs}.
Therefore, they can be removed 
by collinear counter terms generated
by the \lq renormalization\rq   of the incoming parton densities.
The choice of the finite part of the collinear
counter terms is arbitrary 
and allows   the definition of the  different
 factorization schemes.
Of course  the physics  under a change of the
factorization scheme remains the same since
the change in the parton cross-sections
is compensated with the change in the parton densities.
 The collinear
subtraction terms
define  the kernels of the scale evolution of the
parton number densities (Altarelli-Parisi equation~\cite{altpar}).

In the parton model, the  differential
cross section for hadron collisions has the form
\be
d\sigma_{AB}(p_A,p_B)=\sum_{ab}\int dx_1 dx_2 f_{a/A}(x_A,\mu)
f_{b/B}(x_B,\mu)d\hat{\sigma}_{ab}(x_A p_A,x_B p_B,\mu)\,,
\label{factth}
\ee
where $A$ and $B$ are the incoming hadrons, $p_A$ and $p_B$
their momentum, and $a,b$ run over all the parton flavors
which can contribute. 
$d\hat{\sigma}_{ab}$ denotes the finite partonic cross section,
in which the singularities due to collinear emission
of incoming partons are subtracted and
the scale $\mu$ is the factorization scale.

Equation~(\ref{factth}) applies also when 
the incoming hadrons are formally substituted for partons.
In NLO their densities   are singular
\be
f_{a/d}(x)=\delta_{ad}\delta(1-x)-\frac{\as}{2\pi}
\left(\frac{1}{\epb}P_{a/d}(x,0)-K_{a/d}(x)\right)
+{\cal O}\left(\as^2\right),
\ee
where $P_{a/d}(x,0)$ are the Altarelli-Parisi kernels in four 
dimensions ( we  use  $4-2\ep$ dimensions and the $0$ 
in the argument of $P_{a/d}$ stands for $\ep=0$). The finite functions
$K_{a/d}$ are arbitrary, the 
 ${\rm \overline{MS}}$ subtraction scheme 
is obtained by choosing $K_{a/d}\equiv 0$. Expanding
 the unsubtracted and subtracted partonic 
cross sections to  next-to-leading order 
\be
d\sigma_{ab}=d\sigma_{ab}^{(0)}+d\sigma_{ab}^{(1)}\,,\;\;\;\;
d\hat{\sigma}_{ab}=d\hat{\sigma}_{ab}^{(0)}+d\hat{\sigma}_{ab}^{(1)}\,,
\label{decomposition}
\ee
we obtain
\bea
d\hat{\sigma}_{ab}^{(0)}(p_1,p_2)&=&d\sigma_{ab}^{(0)}(p_1,p_2)
\\*
d\hat{\sigma}_{ab}^{(1)}(p_1,p_2)&=
&d\sigma_{ab}^{(1)}(p_1,p_2)
+ d\sigma_{ab}^{\rm count}(p_1,p_2)
\label{counterterms}
\eea
where
\bea
 d\sigma_{ab}^{\rm count}(p_1,p_2)&=&
\frac{\as}{2\pi}\sum_d\int dx\left(\frac{1}{\epb}P_{d/a}(x,0)
-K_{d/a}(x)\right)d\sigma_{db}^{(0)}(xp_1,p_2)
\nonumber \\*&&
+\frac{\as}{2\pi}\sum_d\int dx\left(\frac{1}{\epb}P_{d/b}(x,0)
-K_{d/b}(x)\right)d\sigma_{ad}^{(0)}(p_1,xp_2)\,. \nonumber
\\*
\label{counterterms2}
\eea
Equation~(\ref{counterterms2}) defines the  
 collinear counter terms for any finite hard scattering cross section.
The parton number densities at a given
scale have to be extracted from the data. 
One can systematically 
improve the accuracy of the predictions by calculating
also higher order corrections.
The accuracy of  the  recent experimental
results requires the inclusion of  higher order radiative corrections 
for a large number of measured quantities~\cite{ mangano}.

\begin{figure}[htbp]
\vspace*{-.2cm}
\centerline{ \epsfig{figure=
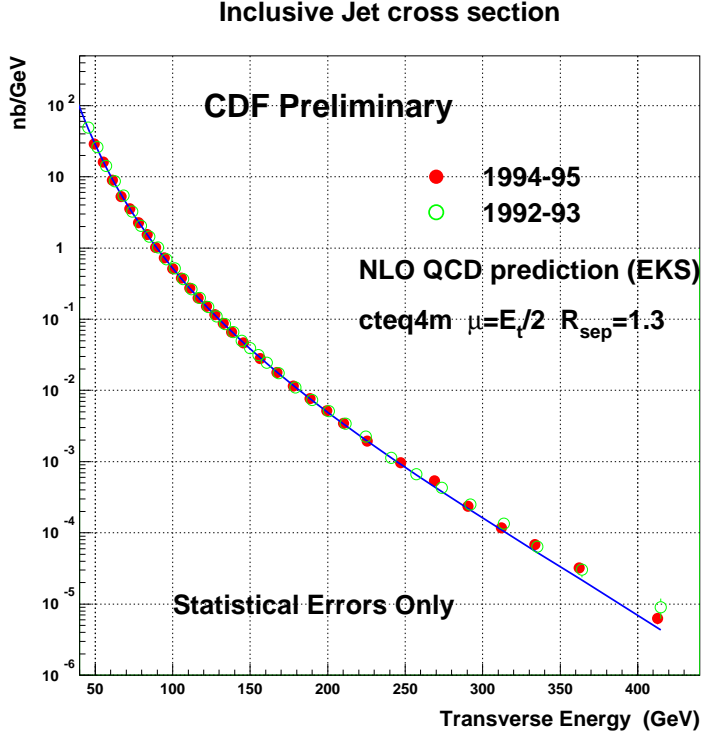,
width=0.7\textwidth,clip=} }
\vspace*{0.21cm}
\caption{\hspace*{0.2cm} Data points obtained by CDF~\cite{abeCDFjet} 
 compared with the NLO
prediction of the theory~\cite{EKS} for the $p_T$ distribution of inclusive
jet production 
at factorization and renormalization scale $\mu=p_t/2$ with
cteq4m parton number densities and the jet separation parameter
$R_{\rm sep}=1.3$. 
}\label{CDFjet}
\end{figure}                                                              


We have a large number of quantitative well tested prediction for
deep inelastic scattering, $W/Z$ production, jet-production and
heavy quark production. The method of calculation
follows of the procedure described above. Both the theoretical
and experimental error could be significantly 
 reduced during the last decade
 and bt now they are about $15\%-30\%$. As illustration of this 
in Fig.~\ref{CDFjet} the inclusive transverse energy $E_T$ distribution
for jet production at Tevatron 
is shown. The data points obtained by CDF~\cite{abeCDFjet}  are compared by
the absolute NLO QCD prediction~\cite{EKS}.
These data are interesting since i) they test the absolute prediction
of QCD up to NLO accuracy ii) they give information on the parton
number densities at large $Q^2$ and iii) they can be used
to constrain possible new physics. 


\begin{figure}[htbp]
\vspace*{-4cm}
 \begin{minipage}{0.47\linewidth}
   \centering
\vspace*{1cm}
\epsfig{file=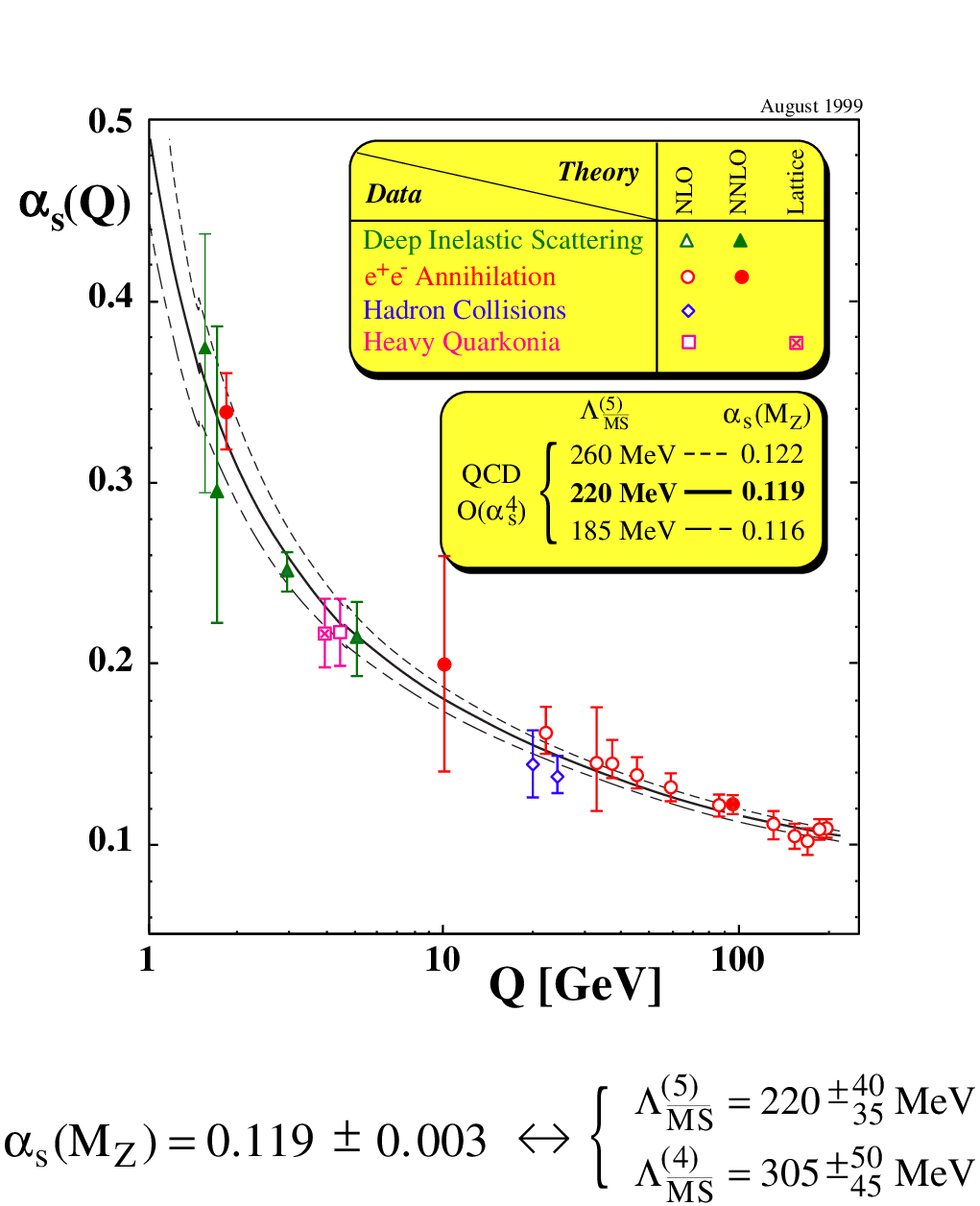,
height=16.5 cm,width=11.5 cm}
\vspace*{-3cm}
   \caption{
World summary of $\as (Q)$ at the energy scales of the experiments
~\cite{bethke}.
} \label{as-q}
 \end{minipage}%
 \begin{minipage}{0.06\linewidth}
   \hspace{0.3cm}
 \end{minipage}%
 \begin{minipage}{0.47\linewidth} 
   \centering
 \epsfig{file=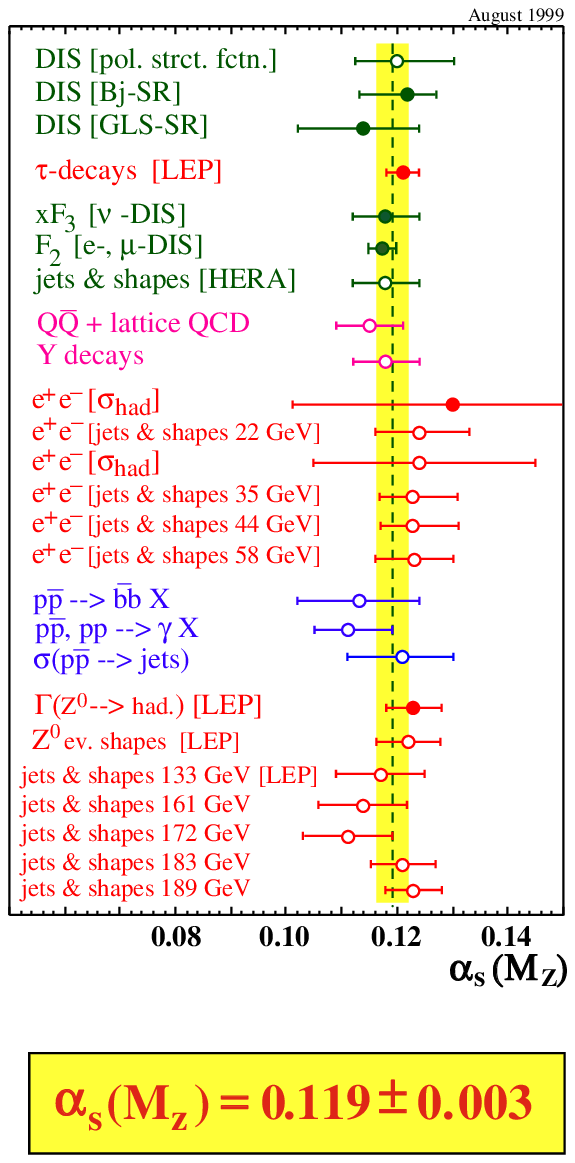,
height=18.5 cm,width=10.5 cm}
\vspace*{-4cm}
   \caption{
World summary of $\as(Q)$ evolved to a common
energy scale $Q=\MZ$~\cite{bethke}.
} \label{as-mz}
 \end{minipage}
\end{figure}

\subsubsection{Measurements of $\as$}
QCD has been discovered by its
 qualitative properties. Over the last 20 years, however, 
significant progress has been achieved in the
field theoretical techniques for deriving its consequences and
by now we can test QCD  ${\it quantitatively}$.
The free parameters of QCD are the quark masses and the  
strong coupling constant. 
Once we fitted these values to some
measurable infrared safe quantities,   
 hard scattering processes  can be predicted from first principles
for $\epem $ annihilation. For processes with hadrons in the initial
state the parton number densities should also be fitted. We do not 
want to go into this detail (see ref.~\cite{ESWbook}).
The status of QCD tests is well characterized with
the agreement  between the various values of $\as$ extracted
from different measurements. 
We illustrate the present situation with two figures~\cite{bethke}.
Fig.~\ref{as-q} gives a test for the running of $\as(Q)$ with plotting
the value of $\as(Q)$ as obtained by various experiments. The data
confirms the energy dependence of $\as(Q)$ as predicted by QCD.
In Fig.~\ref{as-mz} we show the summary of the results of the same
experiments but evolving the values of $\as(Q)$ to a common
energy scale, $Q=\MZ$ using the QCD $\beta$ function in ${\cal O}(\as^4)$
with 3-loop matching at the heavy quark pole masses $M_b=4.7\gev$
and $M_c=1.5\gev$.
 The corresponding world average is also given as well as the
values of $\LMSB$ for $n_f=4,5$.


%
\subsection{Testing  the electroweak theory}

In the Standard Model at tree level  the gauge bosons
$\gamma,W,Z$ and  their interactions 
are described in terms of three parameters:
 the two gauge coupling constants  $g,g^{\prime}$ and
 the vacuum expectation value of the Higgs field $v$.
  We need to know their values
as precisely as possible.
They  have to be fitted 
to  the three  best measured physical quantities
of    smallest 
experimental error: 
 $G_{\mu},\MZ$ and $\alpha$.
In   leading order  we have the simple relations 
\begin{equation}
G_{\mu}=\frac{1}{\sqrt{2}v^2}\,,\quad 
\MZ=\frac{1}{2\cos\theta_W}gv\,,\quad 
\alpha=\frac{g^2}{4\pi}\sin^2\theta_W\,,\quad  
\tan\theta_W=\frac{g^{\prime 2}}{g^2}\,.
\end{equation}
 The muon coupling 
$G_{\mu}$ is extracted from the precise measurement of the muon lifetime
using   the theoretical expression given by equation~(\ref{GF})
\begin{equation}
G_{F}=(1.16637\pm 0.00001)\times 10^{-5}\gev^{-2}\,.
\end{equation}
The  value of $\MZ$
 is extracted  from the line shape measurement at the $Z$-pole.
There are subtleties in the theoretical definition of the mass
and the width at higher order associated with the truncation
of the perturbative series and gauge invariance. The latest best value
is \cite{EWWG2000}
\begin{equation}
\MZ=(91.1871\pm 0.0021)\gev\,.
\end{equation}
The best value of $\alpha$ is extracted 
 from the precise measurement of the electron anomalous
magnetic moment $(g_{\rm e}-2)$~\cite{PDGcaso}
\begin{equation}
{1}/{\alpha }=137.03599959\pm 0.00000038\,.
\end{equation}
Additional  physical quantities
like the mass of the W-boson $\MW$, the lepton asymmetries at the
Z-pole, the leptonic width of the Z-boson $\Gamma_{\rm l}$ 
\etc\  are derived quantities.
At the level of the per mil accuracy the predictions
obtained in Born approximations for derived
quantities, however, disagree with the measured values  significantly.

\subsubsection{Quantum corrections}

 At LEP, SLC and Tevatron  an enormous amount of data has been collected 
 on the  Z and W  bosons
  and their interactions \cite{Swartz:1999xv,Sirlin:1999zc}.
This allows  for  an unprecedented precision test
 of the Standard Model at the level of the per mil accuracy.
At this precision  one
 and two-loop
quantum fluctuations give measurable contributions and  
data data show sensitivity also to the Higgs mass and the 
top mass.

Since the Standard Model
is a renormalizable quantum field theory 
the theoretical 
 predictions of the theory can be improved systematically 
by calculating   higher order corrections.
In particular, the recent  precision of the data
 requires  the study  
of the complete  next-to-leading
order corrections,  resummation of  large logarithmic
contributions  and  a number of 
 two loop corrections. 
 At higher order 
the derived quantities  show sensitivity also to  
the values of the mass parameters  $\mt$, $\MH$, $m_b$ and the QCD
coupling constant
$\alpha_s$.
From direct measurements one obtains 
$\alpha_s=0.119\pm0.002$ (see Fig.~\ref{as-mz}),
 $\mt=173.8\pm 5.0\gev$ and $\bar{m}_b(\bar{m}_b)=4.25\pm 0.08\gev$,
\, $\MH \ge 102\GeV$ where $\bar{m}_b(\bar{m}_b)$ denotes
the $\MSbar$ mass~\cite{Beneke:1999fe}.
The error bars give    parametric
uncertainties in the predictions and  limit our
ability to extract a   precise value of the Higgs mass.
The calculation of the higher orders requires 
 a choice of the renormalization scheme.\footnote{
For a complete discussion of this technical detail with 
references see \cite{BPbook}.}

In the perturbation theory the higher order 
 effects can be given in  bare
and renormalized parameters.
 Let us consider  the basic 
observables  as the basic  bare parameters
 $a_0^i\equiv (G_0,
\alpha_0, M_{Z0})$. Calculating the radiative
corrections in the regularized theory 
the  radiatively corrected renormalized values of the basic parameters
can be written as
\be
a^i(a^i_0)=a_0^i+\delta a^i(a_0^i)\,.
\ee
This relations can be inverted 
\be
a_0^i\equiv a_0^i(a^i)\,.
\ee
Similarly for derived observables we can write
\be
O(a_0^i)\equiv O_0(a_0^i)+\delta O (a_0^i)\,.
\ee
At one loop we can express the radiatively corrected derived quantities
as functions of the radiatively corrected basic quantities
\bea\label{derived1}
O(a_0^i(a_i)) &\approx& O_0(a_0^i)+\delta O (a_0^i)
\nonumber\\
&\approx &O_0(a^i)+\delta^{(1)} O(a^i)-\sum_i\frac{\partial O_0}{\partial a^i}
\delta^{(1)}a^i
\nonumber\\ 
&=& O_0(a_i)+\Delta^{(1)}(a^i)\,.
\eea
The coefficients of this expansion are finite  since
we expand renormalized finite quantities in terms of the
renormalized finite basic observables. Equation~(\ref{derived1}), however,
tells us that the finite one-loop corrections have a direct 
part $\delta^{(1)}O$ and an indirect one 
$-\sum_i\frac{\partial O_0}{\partial a^i}\delta^{(1)}a^i$
coming from  corrections to the basic parameters. The two
contributions can be separately divergent, only the sum is finite.

In the on shell scheme~\cite{BPbook} 
the mixing angle  
$s^2=\sin^2\theta$
 is defined  by  the tree level relation
$s^2=1-\MW^2/\MZ^2$ in all order, therefore, it is a 
 physical quantity. 
It is customary to define auxiliary dimensionless parameters
 $r_W$  by the relation 
\be
s^2c^2\equiv \frac{\pi\alpha}{\sqrt{2}G_{\mu}\MZ^2(1-r_W)}
\ee
where $c^2=1-s^2$. Obviously $r_W$ is also a  finite derived quantity
and it gives the radiative corrections
to the $\MW$.

 In the
 $\overline{\rm MS}$-scheme the measured values of
$\alpha$, $G_{\mu}$, $\MZ$, $m_f$, $\as$
 are used to fix the input parameters of the theory with  $\MH$  as free parameter.
The $\MSbar$ gauge couplings evaluated 
at the scale of $\MZ$ 
are denoted as $\hat{e}$ and $
\hat{s}^2 =\sin^2\hat{\theta}_W(\MZ)$.
The renormalized parameters $\hat{s
}^2$, $\hat{e}^2$ can be completely
calculated  in terms of $G_{\mu}$, $\alpha$ and $\MZ$.
Another useful auxiliary quantity is the effective mixing angle
\be\label{seff}
\sin^2\theta^{eff}_W=\frac{1}{4}\left(1-\frac{\bar{v}_{l}}{\bar{a}_{l}}\right)
=s^2(1+\Delta k^{'}) 
\,.
\ee
This is uniquely related to the ratio of the effective
neutral current vector and axialvector 
couplings  $\bar{v}_l$ and $\bar{a}_l$ of leptons (see equation (\ref{afvf}))
and it gives
the leptonic forward-backward 
 asymmetries $A_{FB}^l$ at the Z-pole in all order of the  perturbation theory
as well as the  tau polarization $A_{\tau}^{\rm pol}$
 and left-right polarization
asymmetries $A_{\rm pol}$
\be
A_{\rm FB}^l=\frac{3}{4}A_eA_f\,,\quad A^{\rm pol}=A_{\tau}\,,
\quad A_{\rm LR}=A_e\,
\ee
where
\be
A_f=\frac{2\bar{v}_v\bar{a}_f}{\bar{v}_f^2+\bar{a}_f^2}\,.
\ee
Measurements of the asymmetries hence are measurements
of the effective mixing angle and therefore
it is a physical quantity as well 
 the dimensionless parameter $\Delta k^{'}$ defined by equation (\ref{seff}). 
The leptonic width depends on the vector and  axial vector
coupling and on the  corrections to the
$Z$-propagator. This requires the introduction of the so called
 $\rho$-parameter
\be
\Gamma_l=
\frac{G_{\mu}\MZ^3 }
{6\pi\sqrt{2}}\left(\bar{g}_{Vl}^2+\bar{g}_{Al}^2\right)
\rho \,.
\ee
The corrections $\Delta r_W$,  $\Delta k'$, $\Delta \rho$  
 are known in various 
schemes and play an important role in the analysis of
electroweak physics, because they give  the 
precise predictions of the theory for simple observables as 
$\MW$, the leptonic
asymmetries \etc\  in terms of $\alpha,G_{\mu}$ and $\MZ$.
It is very useful to have the results 
in different schemes since it allows for cross-checking
the correctness of the result  and to  estimate
 the remaining theoretical errors given by the missing higher order
contributions.  

In the precision tests assuming that  
 the analysis is not restricted to the
Standard Model the radiative corrections it is  convenient
to use the  $\ep$ parameters~\cite{altbar} defined as
\be
\ep_1=\Delta \rho\,,\quad
\ep_2=c^2\Delta \rho+\frac{s^2 \Delta r_W}{(c^2-s^2)}\,,\quad
\ep_3=c^2\Delta \rho + (s^2-c^2)\Delta k^{'}\,.
\ee  
The electroweak radiative corrections are dominated by two leading
contributions: the running of the electromagnetic
coupling and large  
$m_t$ effects to $\Delta\rho$
$( \Delta \rho_t\approx 3G_{\mu}\mt^2/(8\pi^2\sqrt{2}))$.
These corrections can be absorbed into the parameters
of the Born cross section when we get improved Born approximation.

\subsubsection{Running electromagnetic coupling}

The running of $\alpha$ is completely
given by the photon self energy contributions
\begin{equation}
\alpha(\MZ)=\frac{\alpha}{1-\Delta\alpha}
\end{equation}
where
\begin{equation}
\Delta\alpha=
-{\rm Re}\left({\hat\Pi}^{\gamma}(\MZ^2)\right)=
-{\rm Re}\left({\Pi}^{\gamma}(\MZ^2)\right)+
{\rm Re}\left({\Pi}^{\gamma}(0)\right)\,.
\end{equation}
The self energy contribution is large
($\approx 6\%$). It can be split into leptonic
and hadronic contributions
\begin{equation}
\Delta\alpha=\Delta\alpha_{\rm lept}
+\Delta\alpha_{\rm had}
\end{equation}
The leptonic part is known up to three loop \cite{steinhauser} 
\begin{equation}
\Delta\alpha_{\rm lept}=314.97687(16)\times 10^{-4}
\end{equation}
and the remaining theoretical error is completely negligible.
The hadronic contribution is more problematic as it  
can not be calculated theoretically with the required precision
since the light quark loop contributions have non-perturbative
QCD effects. One can extract it, however, from the data using the
relation
\bea
\Delta\alpha_{\rm had}&=&\frac{\alpha}{3\pi}\MZ^2{\rm Re}
\int_{4m^2_{\pi}}^{\infty}ds
\frac{R_{\epem}(s^{\prime)}}{s^{\prime}(
s^{\prime}-\MZ^2-i\epsilon)}\nonumber\\
R_{\epem}(s)&=&\frac{
\sigma(
\epem\to\ \gamma^*\to \ {\rm hadrons})}
{\sigma(
\epem\to\ \mupmum)}\,.
\eea
Conservatively,  one calculates the high energy $\sqrt{s}\ge 40\gev$
contribution using perturbative QCD. The low energy contribution
 $\sqrt{s}\le 40\gev$
is estimated using data \cite{Jegerlehner:1999hg}.
 Unfortunately, the precision 
of the low energy data is not good enough and the error  from this
source  dominates the error of the theoretical predictions
\be
\Delta\alpha_{\rm had}=0.02804\pm0.00064\,,\\
\alpha^{-1}(\MZ)=128.89\pm0.09\,.
\ee
One can, however,  achieve a factor of three
reduction of the estimated error
assuming that the theory can be used 
down to $\sqrt{s}=m_{\tau}$ when
quark mass effects can be included up to three loops.
  Such an analysis
is quite well motivated  by  the successful
results on the   tau lifetime. In the hadronic vacuum polarization
 the non-perturbative power corrections appear to be  suppressed
and the unknown  higher order perturbative
contributions are relatively small. In this theory
driven approach the error is reduced 
to an acceptable $0.25\% $ value
\be
\alpha^{-1}(\MZ)=128.905\pm0.036\,.
\ee
It is unlikely that the low energy hadronic total 
cross section will be measured in the
foreseeable future 
with a precision leading to essential improvement.

\subsubsection{Calculation of $\Delta\rho_t$}

We have noted in section 1.26 that in the limit
of custodial symmetry in leading order $\rho=1$ and the 
dominant radiative correction comes from the virtual effects
of the top quark since the top-bottom mass splitting gives
the largest violation of custodial symmetry. The importance of this
correction was first pointed out by Veltman~\cite{veltman1977}.
It is an elegant technical trick to calculate this 
correction using the effective field theory obtained
in the $m_t\to \infty$ limit~\cite{barbieri-efftop}.
In this limit we need to keep only
the third generation and the gauge bosons can be treated
as external classical currents without kinetic terms.
The Standard Model Lagrangian can be reduced to the terms
\be
{\cal L}_{\rm eff}=
i\overline{\psi}^Q_L\gamma_{\mu}D^{\mu}\psi^Q_L +
i\overline{\psi}^t_R\gamma_{\mu}D^{\mu}\psi^t_R +
i\overline{\psi}^b_R\gamma_{\mu}D^{\mu}\psi^b_R 
+ \lambda_t\overline{\psi}^Q_L\Phi\psi^t_r - V(\Phi)  
\ee
with the 
Higgs doublet  
\bea
 \Phi=\left( \begin{array}{c}
\phi^+ \\ \frac{1}{\sqrt{2}}(v+h+i\chi) \end{array} \right) 
\eea
where $\chi$ and $\phi^+$ are the Goldstone bosons.
The renormalized Lagrangian then has the form
\be
{\cal L}_{\rm eff}=
Z_2^{\phi}\left|\partial_{\mu}\phi^+ - i\frac{gv}{2}
W^+_{\mu}\right|^2 + \frac{Z_2}{2}^{\chi}
\left|\partial_{\mu}\chi-i\frac{gv}{2c}
Z_{\mu}\right|^2 + \frac{Z_h}{2}\left(\partial_{\mu}h\right)^2+...
\ee
where we dropped the top and bottom kinetic energy terms,
the top mass terms and the gauge boson fermion couplings.
In
this limit the gauge boson couplings do not get corrections
but the $Z$ and $W$ mass terms are modified by 
the self energy corrections of the Goldstone bosons
\be
\MW^2=Z_2^{\phi}\frac{g^2v^2}{4}\,, \quad 
\MZ^2=Z_2^{\chi}\frac{g^2v^2}{4c^2}
\ee
and therefore
the correction to the $\rho$ parameter is
\be
\Delta \rho=\frac{Z_2^{\phi}}{Z_2^{\chi}}-1
\ee
that is we can get  the correction
to the $\rho$ parameter simply by calculating the
the contributions of top and top-bottom fermion loops to
difference of the  self energies of the neutral
and charged Goldstone bosons.
Carrying out this simple
calculation we can easily check that the answer is
\be
\Delta \rho_t \approx 3\frac{G_Fm_t^2}{8\pi\sqrt{2}}.
\ee
The method can be extended  to two loop order 
and the corresponding two loop calculation has been carried out
in ~\cite{barbieri-efftop} confirming previous result~\cite{vanderbij}.
\begin{figure}[htbp]
\vspace*{-1cm}
\centerline{ \epsfig{figure=
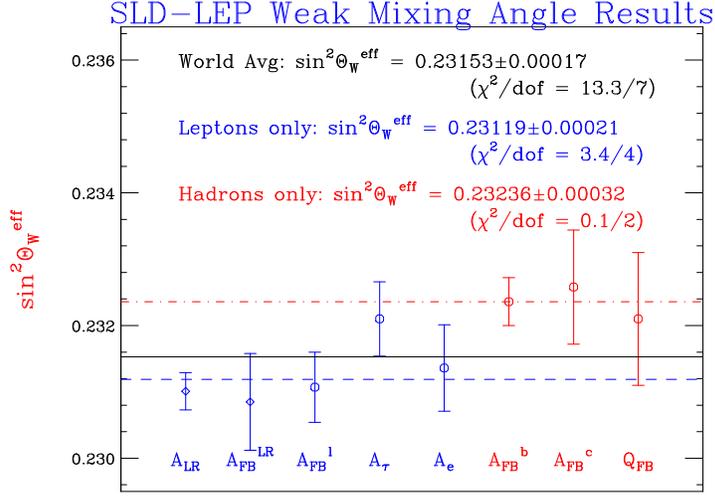,
width=0.7\textwidth,clip=} }
\vspace*{-1cm}
\caption{\hspace*{1cm}Summary of all the determinations of 
  $\sin^2\theta_W^{\rm eff}$ \cite{Swartz:1999xv}.
}\label{seff-fig}
\end{figure}                                                              

\subsubsection{Higher order corrections to $\MW$ and the mixing angle }

As we noted above, the simplest physical observables for  
precise tests of the Standard Model are
$\MW$ and the $\sin^2\theta^{eff}_W$. 
It is convenient to consider the radiative corrections
in the $\MSbar$ scheme where with good accuracy
$\sin^2\theta^{eff}_W\approx \hat{s}^2$.
It is given in terms of the input parameters
via the relation
\be
\hat{s}^2\hat{c}^2=\frac{\pi\alpha(\MZ)}{\sqrt{2}G_{\mu}\MZ^2(1-\hat{r}_W)}
\ee
where  $\hat{r}_W=0$ in leading order. Using the measured value of
$\sin^2\theta^{eff}_W,\MZ$ and $G_{\mu}$ we obtain a value
$\hat{r}_{W}=0.0058\pm 0.000480$  different from zero at the
$12\sigma$
level. If one carries out a similar analysis for $\MW$ the evidence for
the presence of subleading corrections  is even better.
 The radiative correction 
$\hat{r}_W$ does not contain the large
effect from the running $\alpha$ but it receives  large
custodial symmetry  violating corrections because of  the large
top-bottom mass splitting
\be
\Delta \hat{r}_W|_{\rm top}
=-c^2/s^2\Delta \rho \approx  
0.0096  \pm 0.00095 \,.
\ee
Subtracting this value we get   about $6\sigma$ difference coming
from   the loops involving the bosonic
sector (W,Z,H) and subleading fermionic contributions. 
At this level of
accuracy many other corrections start to become important and  the
 size of  errors coming from the errors in the input parameters
leads to effects of the same order. In particular, we get some
sensitivity to 
the value of the Higgs mass. 
 Beyond the complete one
loop corrections it was possible to evaluate    
all  important two loop corrections: 
  ${\cal O}(\alpha^2\ln(\MZ/\Mf)$ corrections 
 with light fermions, mixed electroweak QCD corrections
of ${\cal O}(\alpha\alpha_s)$,
 two loop electroweak corrections
 enhanced by top mass effects of
  ${\cal O}(\alpha^2(\mt^2/\MW^2)^2)$ together with 
 the subleading parts of  
 ${\cal O}(\alpha\alpha_s^2 \mt^2/\MW^2)$ and
the very difficult subleading correction
 of  ${\cal O}(\alpha^2 \mt^2/\MW^2)$.
 It is remarkable that this last  contribution proved to be
 important in several respect \cite{Degrassi:1999jd}.
Its inclusion reduced significantly
 the scheme dependence of the results and lead to a  significant  
 reduction of the upper limit on the Higgs mass.

\subsubsection{Global fits}
 This summer the LEP experiments and SLD could finalize their results
 on the electroweak precision data. The most important development
 is that the final value of SLD on the leptonic polarization asymmetry 
   which implies $\sin^2\theta^{eff}_W=0.23119\pm 0.00020$.
 A nice summary of the results is given in Fig.~\ref{seff-fig}.
According to a recent analysis of the EWWW working group~\cite{EWWG2000},  
the new world average is
\be
\sin^2\theta^{eff}_W=0.23151\pm 0.00017 {\rm \ \ \ with\ \ \ }
 \chi^2/{\rm d.o.f=13.3/7} \,.
\ee
 This gives only rather low  confidence level of $6.4\%$.
 The origin of this unsatisfactory result is the $2.9\sigma$ discrepancy
 between the values  $\sin^2\theta^{eff}_W$ deriving from the SLAC 
 leptonic polarization asymmetry data and from the forward backward asymmetry
 in the b-b channel at LEP and SLC.
 The results obtained from a
 global fit to all data give somewhat better result but there we are hampered
 with the problem that the polarization asymmetry parameters
 disagree with each other with $2.7\sigma$, therefore the $\chi^2$ is
 relatively large.

\begin{figure}[htbp]
\vspace*{0cm}
\centerline{ \epsfig{figure=
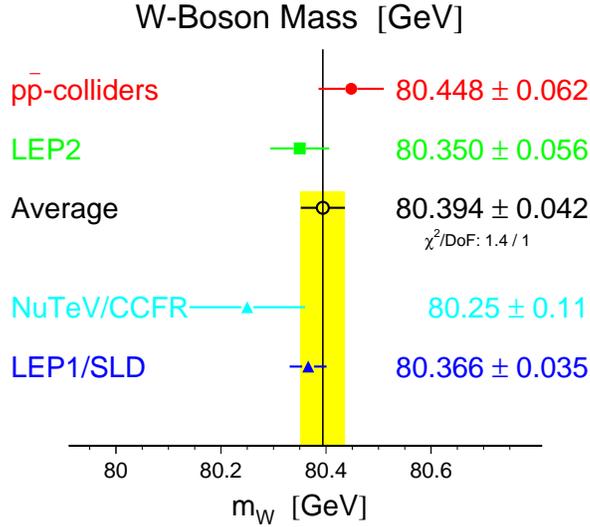,
width=.50\textwidth,clip=} }
\vspace*{0cm}
\caption{\hspace*{1cm}
Summary of the measured values of $\MW$~\cite{Swartz:1999xv}.
}\label{summary-mw}
\end{figure}                                                              

\subsubsection{Upper limit on \MH}
 The final  results of the electroweak radiation
 corrections for $\MW$ and $\sin^2\theta^{eff}_W$
 can be parameterized 
  in terms of the input parameters
 including their errors in simple approximate analytic form 
\cite{Degrassi:1999jd}.
 For example in the $\overline{MS}$-scheme one obtains for the W-mass
\bea\label{MWcorrfit}
\MW&=&80.3827-0.0579\ln (\frac{\MH}{100})
           -0.008\ln^2 (\frac{\MH}{100})\nonumber\\
          && -0.517\left(\frac{\Delta\alpha_h^{(5)}}{0.0280}-1\right)
           +0.543\left[\left(\frac{\mt^2}{175}-1\right)\right]\nonumber\\
          && -0.085 \left(\frac{\alpha_s(\MZ)}{0.118}-1\right)
\eea
where $\mt$, $\MH$ and $\MW$  are in $\gev$ units.
 This formula accurately reproduces the result obtained with 
 numerical evaluation of all corrections in the range 
 $75\gev \le \MH \le 350 \gev$ with  maximum deviation of less than
$1\mev$. In Fig.~\ref{summary-mw} the measured values of $\MW$
are summarized~\cite{Swartz:1999xv}.
 Using  the world average
 $\MW=80.394\pm  0.042\gev$
with input parameters $\alpha_s=0.119\pm 0.003$, $\mt=174.3\pm 5.1\gev$,
$\delta\alpha^{(5)}=0.02804\pm 0.00065$
 one  obtains at $95\%$ confidence level an allowed range for
the Higgs mass of $73\gev \le \MH\le 294\gev$.
 Similar result exists also for $\sin^2\theta^{ eff}_W$ extracted from
 the asymmetry measurements 
at the Z-pole with somewhat better (95\% confidence) limits
 of $95\gev \le \MH\le 260\gev$.
 Without global fits we got a semi-analytic insight on  the sensitivity 
 of the precision tests to the
Higgs mass. We also see that the precise measurements 
of $\MW$
have already provided us with
 competitive values  in comparison with  those obtained
from the measurement of  $\sin^2\theta^{ eff}_W$.

It is interesting that the values of the Higgs mass obtained
in a recent global fit \cite{D'Agostini:2000ws}
 are in good agreement  with the simple analysis based on the value of
 $\MW$ or $\sin^2\theta^{eff}_W$  as described above.
 One obtains an expected value
for the Higgs boson of $160-170\gev$ with error of $ \pm 50-60 \gev$.
The 95\% confidence level upper limit is about
$260-290\gev$. In Fig.~\ref{blueband} $\Delta \chi^2$
is plotted as the function of $\MH$.

\begin{figure}[htbp]
\vspace*{-0.5cm}
\centerline{ \epsfig{figure=
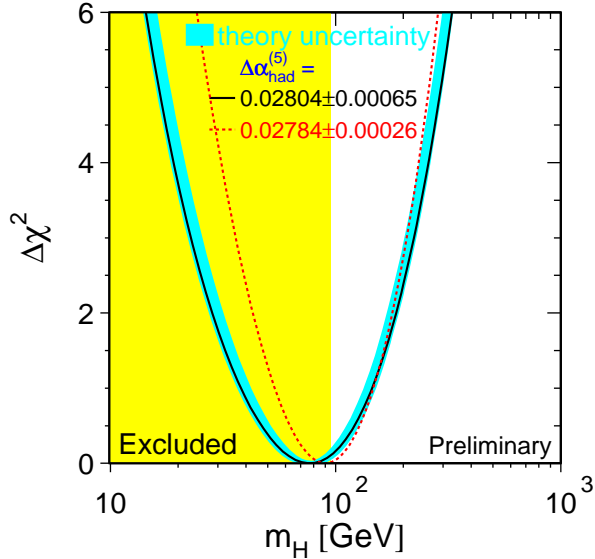,
width=.50\textwidth,clip=} }
\vspace*{-1cm}
\caption{
Significance of the global fit as a function of the 
Higgs mass~\cite{EWWG2000}.
}\label{blueband}
\end{figure}                                                              

\subsubsection{Can the Higgs boson be heavy?}

The precision data can not yet rule out  
 dynamical symmetry breaking with some heavy Higgs like
scalar and vector resonances.
The minimal model to describe this alternative  is obtained
by assuming that the new particles are heavy (more than 0.5 \tev)
 and the linear $\sigma$-model Higgs sector
of the Standard Model is replaced 
 by the non-renormalizable
non-linear $\sigma$-model. It  can  be derived  also as
 an effective chiral vector-boson Lagrangian with non-linear 
realization of the gauge symmetry \cite{ApBe80, longhitano:81}.
How can we reconcile this more phenomenological
approach  with the precision
data?
Removing the
Higgs boson from the Standard Model while keeping
the gauge invariance is a  relatively mild
change.  Although the model
becomes non-renormalizable,  at the one-loop
level  the radiative effects grow only logarithmically with the cut-off
at which new interactions should appear.
In equation (\ref{MWcorrfit}) the Higgs mass is replaced by
this  cut-off.
 The logarithmic terms
are universal, therefore their coefficients
must remain the same. The  constant terms, however, can be different
from those of the Standard Model. The one loop
corrections of the effective  theory
require the  introduction of  new free parameters which
influence the value of the constant terms. 
The data, unfortunately, do not have sufficient precision
to significantly  constrain  
the constant terms appearing in $M_W$, $\sin^2\theta^{eff}_W$
and $\Gamma_l$ (or alternatively in the parameters 
 $\epsilon_1,\epsilon_2,\epsilon_3$~\cite{altbar} or $S,T,U$~\cite{pestak} ).
In a recent analysis ~\cite{Bagger:1999te} 
it has been
 found that due to the screening
of the symmetry breaking sector \cite{veltman},
 alternative theories with dynamical 
symmetry breaking and heavy scalar and vector bosons 
still can be in 
 agreement  with  the precision data 
up to a cut-off scale of $3\TeV$.

\begin{figure}[htbp]
\vspace*{-1.5cm}
\centerline{ \epsfig{figure=
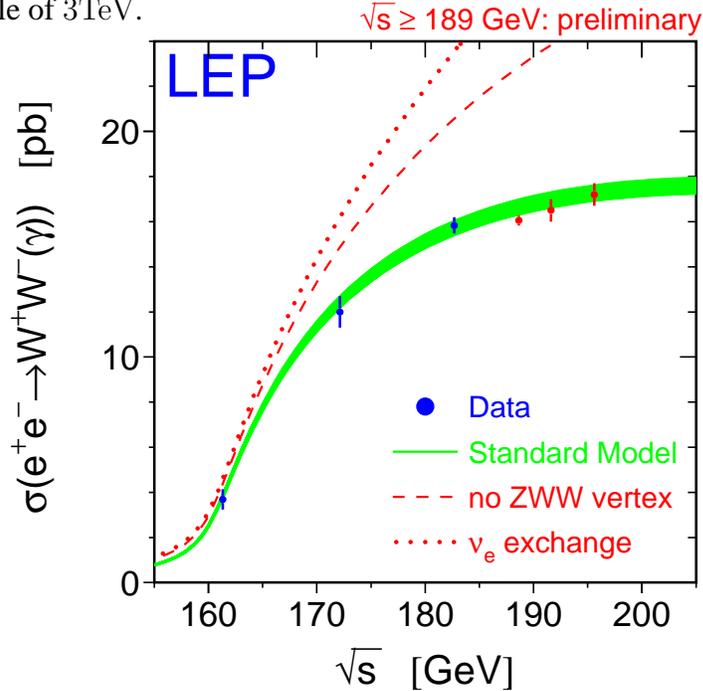,
width=.60\textwidth,clip=} }
\vspace*{-1cm}
\caption{
The W-pair production cross-section as function of the center-of-mass
energy~\cite{EWWG2000}. The data points are the LEP averages. Also shown is
the Standard Model prediction~\cite{bardin} (shaded area)
with a possible uncertainties of $\pm2\%$ of the calculations.
For comparison the cross sections are also shown
if the ZWW coupling did not exist (dashed line), or
if only the t-channel $\nu_e$ exchange diagram existed (dotted line).
}\label{WWLEP}
\end{figure}                                                              

\subsubsection{W-pair production}
At LEP
 the precise measurement of the production
of  $W^+ W^-$
is also an  important physics goal.
The production of gauge boson pairs provide us with the best
test of the non-Abelian gauge symmetry of the Standard Model.
 Deviation from the Standard Model
predictions may come either from the presence of anomalous 
couplings or the production of new heavy particles
and their decays into vector-boson pairs.
If the particle spectrum of the Standard Model has to be enlarged with
new particles (as in the Minimal Supersymmetric Standard Model)
with mass values of $\ge 0.5-1 \TeV$, small anomalous
couplings are generated at low energy. 
In Fig.~\ref{WWLEP} we show the recent measurement of
the W-pair cross-section.

\section{HIGGS SECTOR, HIGGS SEARCH}

\def\Ord{\buildrel{\scriptscriptstyle <}\over{\scriptscriptstyle\sim}}
\def\OOrd{\buildrel{\scriptscriptstyle >}\over{\scriptscriptstyle\sim}}

\subsection{Difficulties with the Higgs sector}
\noindent
The Standard Model is defined only in perturbation theory, but 
the perturbative treatment of the Higgs sector 
can not be  valid up to   arbitrary
high energies. 
Its  range of validity   depends strongly on the  value of the Higgs mass.

 \begin{figure}[hbtp]
\vspace*{0cm}
\centerline{
 \epsfig{figure=
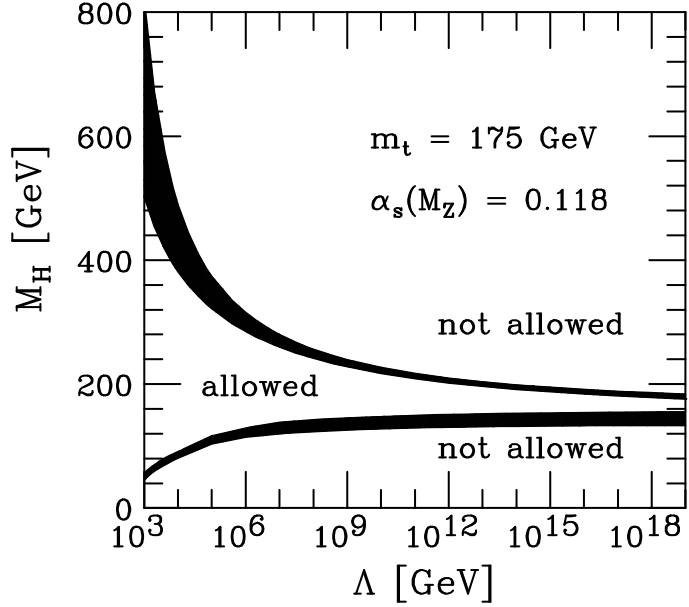, 
width=0.8\textwidth,clip=} }
\vspace*{-5.5cm}
\caption{Bounds on $M_H$ as as a function of the cut-off scale $\Lambda$.
The upper solid area indicates the the triviality upper bound.
The width of the area indicates the sum of the theoretical
uncertainties in the $\MH$.
The lower solid area represents the lower bounds derived from
stability requirements using $m_t=175\gev$ and $\as=0.118$~\cite{altisi,quiros}.}\label{fig:riesselmann}
\end{figure}

\subsubsection{Theoretical upper limits on $\MH$ }
For high Higgs mass values 
we get conflict with perturbative unitarity~\cite{LeeQuigg} 
since according to equation~(\ref{selfcp}) if $\MH>>\MW$ the scalar self
 interaction
becomes strong.
Unitarity requires that in a given angular momentum channel
the scattering matrix element fulfills the relations
\be
\abs{M_J}^2\le \abs{{\rm Im}(M_J)}\,, \quad \abs{{\rm Re}(M_J)}\,<\frac{1}{2}
\ee
Applying these constraints for the Born amplitude of $W_LW_L$ scattering we get
\be
\MH^2\le \frac{2\pi\sqrt{2}}{G_F}\approx (850\gev)^2
\ee
A more precise coupling channel analysis  
leads to somewhat better limit
\be
\MH^2\le \frac{2\pi\sqrt{2}}{G_F}\approx (700\gev)^2\,.
\ee
If the Higgs self coupling is large,   
 the gauge interactions are negligible. The scalar interaction, however,
is not asymptotically free, 
therefore in the perturbation theory 
the running
coupling constant has a Landau pole. Actually it has been proven
that  the scalar 
theory is trivial~\cite{Frohlich}.
If we require to have finite  scalar coupling at very short distances 
we get vanishing coupling at large distances: the theory becomes free.
The scalar sector mathematically
can not be rigorous and it should
be considered as an effective low energy theory.
The scalar self coupling $\lambda$, therefore, has to be smaller than
its value at the Landau pole. This condition
also gives  an upper limit
on $\MH$.
The one loop running coupling is
\be
\lambda(\mu)=\frac{\lambda(\MH)}{1-12\frac{\lambda(\MH)}{16\pi^2}\ln \frac{\mu^2}{\MH^2}}
\ee
where $\lambda (\MH)=g^2\MH^2/(8\MW^2)$. 
The position of the Landau pole  is at the scale
\be
\mu_c=\MH e^{\frac{2\pi^2}{3\lambda (\MH)}}=\MH e^{\frac{16\pi^2\MW^2}{3 g^2\MH^2}}
\ee
the condition $\mu_c > 2\MH$ leads to the upper limit
\be
M_H<700\GeV \,.
\ee
If the Landau pole is pushed up to the Planck scale 
we get the  more stringent limit
of $\MH<170\gev$. 
This is a tentative estimate since  perturbation theory is used
beyond   its range of validity. Note that non-perturbative 
lattice studies~\cite{kuti}
give very similar value $\MH<650\gev$.
One can redo the analysis at two loop order when 
the two loop beta function has a metastable fixed point.
With the assumption that the theory is meaningful up to
a scale where the coupling constant is half of the value of
the coupling at the metastable fixed point one,  gets again 
 similar upper limit~\cite{altisi,quiros,hambye}.

\subsubsection{ Lower limits on $\MH$ }
The requirement of  stability of the Higgs potential $V(\phi)>0$ 
leads to a lower limit on the Higgs mass.
The $\beta$-function of the scalar self-interaction is
\bea
\beta_{\lambda}&=&\frac{1}{8\pi^2}\bigl[12\lambda^2-3 g_t^4
+6\lambda g_t^2 - \frac{\lambda}{2}(9g^{'2}+3g^2  )
\nonumber \\
&&+\frac{3}{16}(3g^{'4}+2g^{'2}g^2+g^4)\bigr]
\eea
where $g_t=g m_t/(2\MW)$ denotes the top quark Yukawa coupling.
At large  top mass and small $\lambda$ 
the second term will dominate, as  we get  
negative $\beta$ function and $\lambda$ will decrease with increasing top mass 
and $V(\Phi)$ can become negative.
 A coupled channel analysis at two loop order
leads to the approximate relation (assuming that the theory
is meaningful up to the Planck scale)
$\MH>1.95 m_t-190\gev$ and for $m_t=175\gev$
one gets  the  lower limit $\MH>150\gev$. So the Higgs boson should not
be found at LEP2 in this case.
In Fig.~\ref{fig:riesselmann} theoretical upper and lower bounds
on $M_H$ are shown as a function of the scale charaterizing
the range of validity (cut-off scale) of the Standard Model.
We can see that if $\MH\approx 165-195 \gev$ the 
cut-off scale can extend up to the Planck scale.
I should recall, however, that this is rather unlikely since
the mass term of the scalar boson is a relevant operator, therefore, 
it is linearly sensitive of the scale of new physics. The question
 ``why is the Higgs mass so light with respect to the Planck scale'' 
is referred to in  literature as the  gauge hierarchy problem.

\subsection{Search at LHC}
One of the most important physics goal at the LHC is  to obtain
decisive experimental test on the Higgs sector
 of the Standard Model~\cite{guide,LHC}.
The experimental prospects are summarized with the so called
``{\it no loose scenario}''
\begin{itemize}
\item[i)]
either the Higgs boson  will be discovered at LEP2 or LEP2 will establish
a lower limit of $m_H>107\gev$;
\item[ii)]
assuming the validity of the minimal Higgs sector
the precision data obtained at  LEP1 and SLC
constrain the value of  $\MH$ to be less than $\approx 295\gev$.
\item[iii)] 
LHC will be able to discover the Standard Model Higgs boson in the
 interval $800\gev > m_H > 107\gev $
or will find  clear evidence for deviations from 
Standard Model predictions. 
\end{itemize}
To be able to make maximal use of the  results of the LHC experiments
 one should calculate the Standard Model
predictions for LHC processes as precisely as possible.
It is unsatisfactory that the present experimental simulation
studies are based on leading order cross sections. Nevertheless at the
parton level most of the NLO corrections are available and are public in form
of  program packages~\cite{spirahgludec}.
HDECAY  generates all branching fractions of the
Standard Model Higgs boson and the Higgs bosons of the 
Minimal Supersymmetric Standard Model (MSSM), while HGLUE
 provides the production 
cross sections of  the SM and MSSM Higgs bosons via gluon fusion
including  the NLO QCD corrections.

\subsubsection{Higgs branching ratios}
The branching ratios of the Higgs boson have been 
studied in many papers. 
A useful compilation of the early works on this subject can be found in
Ref.~\cite{guide}, where  the most relevant formulae
for on-shell decays are summarized.
Higher-order corrections
to  most  of the decay processes have  been computed
(for
up-to-date reviews  see Refs.~\cite{corrreview} 
and references therein).

The bulk of the QCD corrections to $H\to q\bar q$ can 
be absorbed into a `running' quark mass $m_q(\mu)$, evaluated at the energy 
scale $\mu=M_H$. 
The importance of this effect for the case $q=b$, 
with respect to intermediate-mass Higgs searches at the LHC, 
has been discussed already in Ref.~\cite{oldpaper}.
For sake of illustration, 
results on the Higgs branching ratios are summarized 
in Fig.~\ref{fig:brlo}. Branching ratios are depicted as function
of $\MH (  \MH \leq 200~\gev)$
for  channels: (a) $b\bar b$,
$c\bar c$, $\tau^+\tau^-$, $\mu^+\mu^-$ and $gg$; and for 
channels (b) $WW$,
$ZZ$, $\gamma\gamma$ and $Z\gamma$. 
The patterns of the various curves are not significantly different
from those presented in Ref.~\cite{oldpaper}. The
inclusion of the QCD corrections in the quark-loop induced
decays
 give a change of at most a few per cent for the
decays $H\to\gamma\gamma$
and $H\to Z\gamma$, while for $H\to gg$ differences are of order
50--60\%. However, this latter result has little phenomenological relevance,
since this decay width makes a negligible contribution to the total width,
and in practice it is an unobservable channel.

 \begin{figure}[hbtp]
\vspace{0cm}
\centerline{ \epsfig{figure=
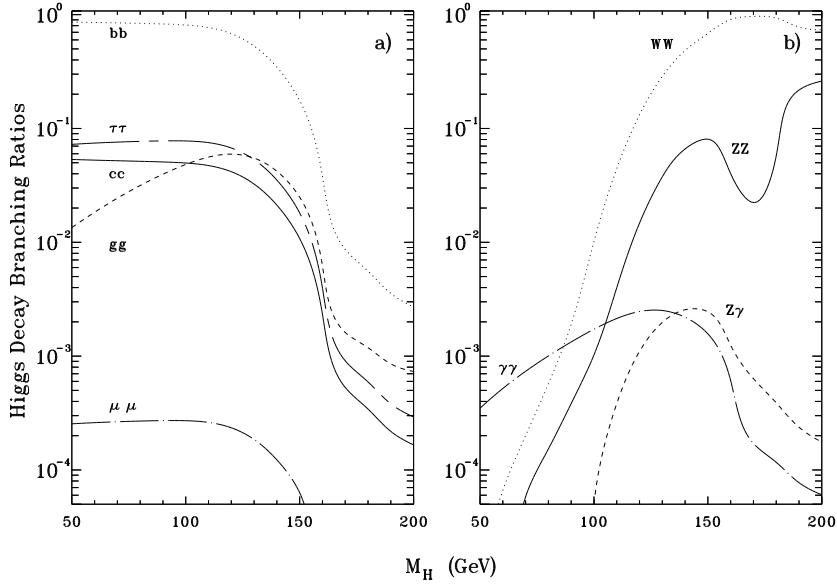,
width=1.0\textwidth,clip=} }
\vspace*{-15cm}
\caption{
Branching ratios of the  Higgs boson in the mass
range 50~GeV $< M_H <$~200 GeV, for the decay modes: a) $b\bar b$,
$c\bar c$, $\tau^+\tau^-$, $\mu^+\mu^-$ and $gg$; b) $WW$,
$ZZ$, $\gamma\gamma$ and $Z\gamma$~\cite{MWJSZK}\,.
}
\label{fig:brlo}
\end{figure}

\begin{figure}[hbtp]
\vspace*{-2cm}
\centerline{ \epsfig{figure=
higgsdiag.ps,
width=0.9\textwidth,clip=} }
\vspace*{-2cm}
\caption{
Feynman diagram for various production mechanisms.
}
\label{fig:diag}
\end{figure}
 \begin{figure}[hbtp]
\vspace*{-8cm}
\centerline{
 \epsfig{figure=
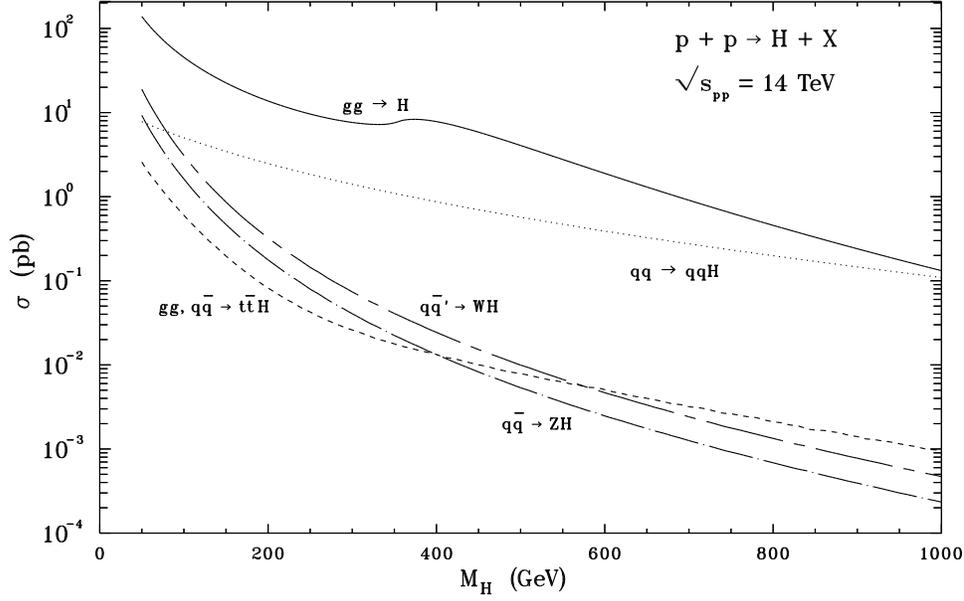,
width=1.00\textwidth,clip=} }
\vspace*{-4.7cm}
\caption{ Total cross sections for $H$ production at the  LHC as a
function of the Higgs mass $M_H$,  given by the four production
mechanisms illustrated in Fig.~\ref{fig:diag}, at $\sqrt s_{pp}=14$~TeV\,.
}
\label{fig:prod}
\end{figure}
 \begin{figure}[hbtp]
\vspace*{-7.5cm}
\centerline{
 \epsfig{figure=
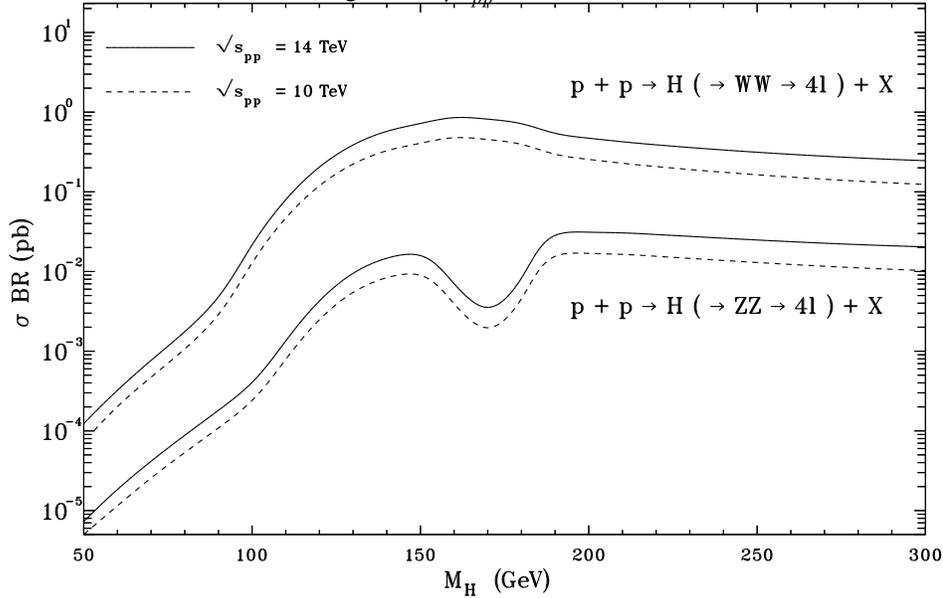,
width=1.00\textwidth,clip=} }
\vspace*{-7.5cm}
\caption{  
Higgs production cross sections times the branching
ratios for the decay modes
 $H\to W^{(*)} W^{(*)} \to 
\ell^+\nu_\ell\ell'^-\bar\nu_{\ell'}$ ($\ell,\ell'=e,\mu$) and
 $H\to Z^{(*)} Z^{(*)} \to 
\ell^+\ell^-\ell'^+\ell'^-\, $
  as a function
of the Higgs mass in the range
  $0 \leq M_H \leq  300$~GeV,
at $\sqrt s_{pp}=10$ TeV and $\sqrt s_{pp}=14$~TeV with
  $m_t=175$ GeV.
}
\label{fig:events}
\end{figure}


 \begin{figure}[hbtp]
\vspace*{-1.7cm}
\centerline{
 \epsfig{figure=
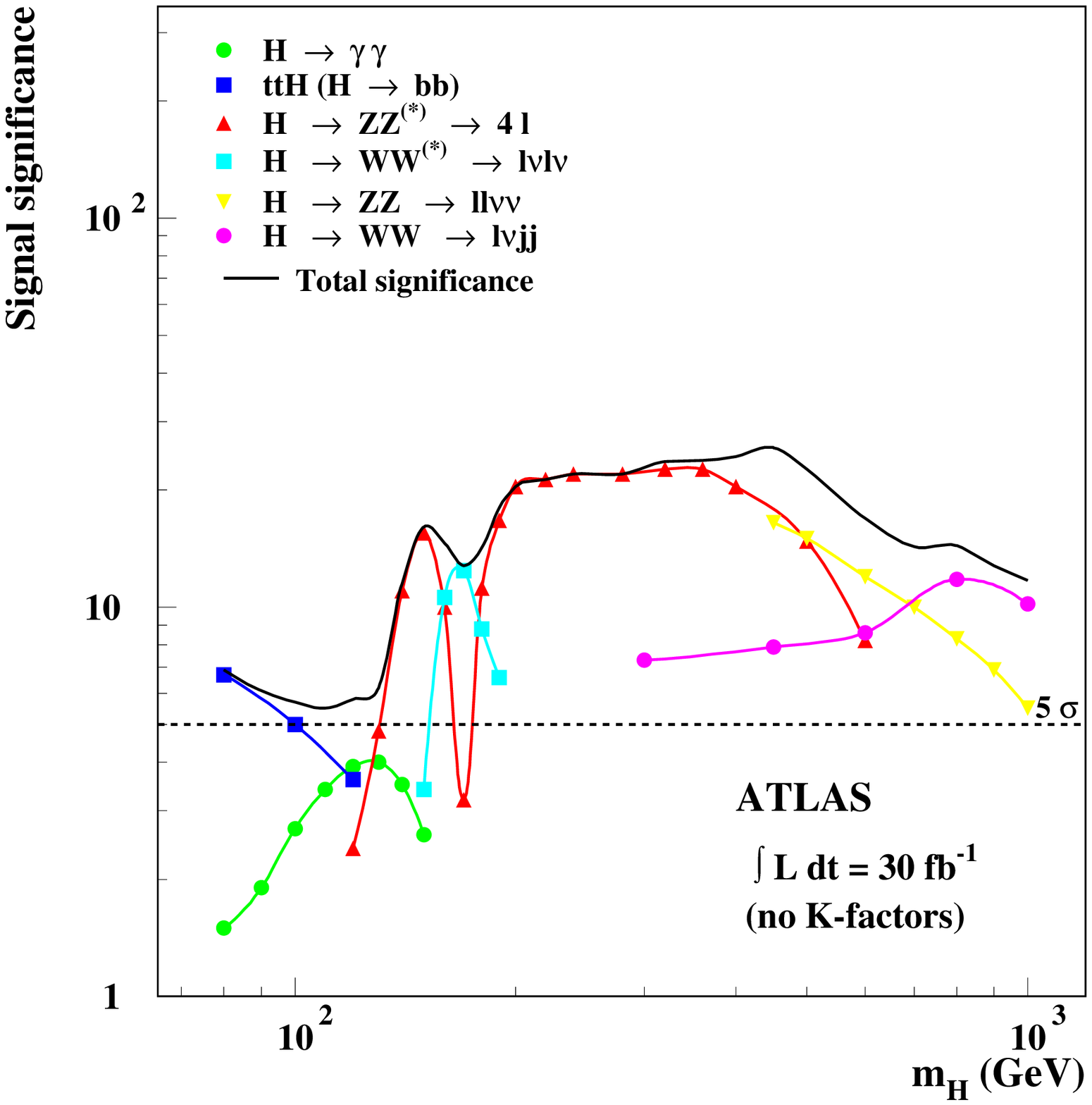, 
width=0.7\textwidth,clip=} }
\vspace*{-0.2cm}
\caption{ Signal significance
as function of the Higgs mass $M_H$ for the ATLAS experiment.
 The statistical 
significances are plotted for individual channels as well
as combination of all channels assuming integrated luminosity
of 30 $\rm fb^{-1}$. Depending on the
numbers of signal (S) and background (B)events, the
statistical significance has been computed as $S/\sqrt{B}$~\cite{atlas-TDR}.
}
\label{fig:atlas1}
\end{figure}

\subsubsection{Higgs production cross sections and event rates}
There are only four Higgs production mechanisms which lead to detectable 
cross sections at the LHC:
a)  gluon-gluon fusion \cite{xggh}, 
b)  $WW$, $ZZ$ fusion \cite{xvvh},
c)  associated production with $W$, $Z$ bosons \cite{xvh},
d)  associated production with $t\bar t$ pairs \cite{xtth}.
 Each mechanism involves heavy particles. 
Representative Feynman diagrams are shown in   Fig.~\ref{fig:diag}.
Again for illustrative purpose in Fig.~\ref{fig:prod}
total cross-section values are depicted 
for LHC energies$\sqrt{s}=14~\tev$ .
There are various uncertainties in the rates of the above 
processes, although none is particularly large. 
They are given by  the lack of precise knowledge  of the gluon
distribution at small $x$
and by effects of
unknown  higher-order perturbative QCD corrections~\cite{MWJSZK}.

The next-to-leading order  QCD corrections  are known
for processes (a), (b) and (c) and are included.
 By far the most important of these
are the corrections to the gluon fusion process 
 calculated in Ref.~\cite{newKfacgg}.
Within the limit  where  the Higgs mass is far below the
$2m_t$ threshold, these corrections are calculable analytically
\cite{Kfacgg2,Kfacgg3}.
In fact, it turns  out that the analytic result is a good approximation
up to the threshold $\MH<2m_t$~\cite{DjouadiRev,sdgz}.
In Ref.~\cite{sdgz,MWJSZK} the impact of the 
 next-to-leading order QCD corrections
for the gluon fusion process on LHC cross sections has been investigated,
both for the SM and for the MSSM.
Overall, the next-to-leading order correction increases 
the leading order result by a factor of about 2, when the normalization
and factorization scales are set equal to  $\mu = M_H$.
This `$K$-factor' can be traced to a large constant piece in the
next-to-leading correction \cite{newKfacgg}
\be
\label{Kfac}
K \approx 1 + {\alpha_s(M_H)\over \pi}\;\left[{\pi^2}+{11\over
2}\right] .
\ee
Such a large  $K$-factor usually implies a non-negligible scale
dependence of the theoretical cross section. 

To judge the quality of the various signals of Higgs production, 
we must know  event rates both for the signals and the backgrounds.
Considering all the possible combinations of
production mechanisms and decay channels~\cite{ATLAS,CMS},
the best chance of discovering a  Higgs at the LHC 
are given by the signatures: 
(i) $gg\to H\to \gamma\gamma$, (ii)
$q\bar q'\to WH\to \ell\nu_\ell\gamma\gamma$ and (iii) $gg\to H\to
Z^{(*)}Z^{(*)}\to \ell^+\ell^-\ell'^{+}\ell'^{-}$, where $\ell,\ell'=e$
or $\mu$.
 By exploiting techniques of
flavor identification of $b$-jets, thereby reducing the huge QCD
background from light-quark and gluon jets, the modes
 (iv) $q\bar q'\to WH\to \ell\nu_\ell b\bar
b$ and (v) $gg,q\bar q\to t\bar t H\to b\bar b b\bar b WW
\to b\bar b b\bar b  \ell\nu_\ell X$, can also be used to search 
for the Higgs \cite{richter}. 
Another potentially important channel, particularly
for the mass range $2 M_W \Ord M_H  \Ord 2 M_Z$,  is
(vi) $ H\to W^{(*)}W^{(*)}\to \ell^+\nu_\ell \ell'^-\bar\nu_{\ell'}$
 \cite{dittdrei}. Here the lack of a measurable narrow resonant
peak is compensated by a relatively large branching ratio,
since for this mass range $H\to WW$ is the dominant decay mode.
Again for sake of illustration we show in Fig.~\ref{fig:events}
 Higgs production
times branching ratios for various decay modes at two different
energies.

The potential of the ATLAS experiment for the discovery
of the Standard Model Higgs boson in the mass range
$80\gev <\MH< 1\tev$ is summarized in Fig.~\ref{fig:atlas1}.
The significance of the signal depends on the signal (S)
and background events and is given by $S/\sqrt{B}$.
The results shown in the figure are obtained from
calculating the event rates both for the
background and the signal in the Born approximation.
It is argued that since the QCD corrections are not
known for all signal and background processes,     
it is more consistent to neglect them everywhere.
Hopefully this shortcomings will be eliminated soon.
One can consider these result conservative since
the QCD corrections are large for the signal
and there is no reason to assume that they
are even larger for the background.



\begin{thebibliography}{999}
\parskip 0pt
\itemsep=0pt

\def\ap#1#2#3{{   Ann. Phys. (NY) }{\bf #1} #2 #3}
\def\app#1#2#3{{  Acta Phys. Pol. }{\bf #1} #2 #3 }
\def\ar#1#2#3{{   Ann. Rev. Nucl. Part. Sci. }{\bf #1} #2 #3 }
\def\cmp#1#2#3{{  Commun. Math. Phys. }{\bf #1} #2 #3 }
\def\cpc#1#2#3{{  Comput. Phys. Commun. }{\bf #1} #2 #3 }
\def\ijmp#1#2#3{{ Int .J. Mod. Phys. }{\bf #1} #2 #3 }
\def\ibid#1#2#3{{ ibid }{\bf #1} #2 #3 }
\def\jmp#1#2#3{{  J. Math. Phys. }{\bf #1} #2 #3 }
\def\jetp#1#2#3{{ JETP Sov. Phys. }{\bf #1} #2 #3 }
\def\jetpl#1#2#3{{ JETP Letter }{\bf #1} #2 #3 }
\def\mpl#1#2#3{{  Mod. Phys. Lett. }{\bf #1} #2 #3 }
\def\nat#1#2#3{{  Nature (London) }{\bf #1} #2 #3 }
\def\nc#1#2#3{{   Nuovo. Cim. }{\bf #1} #2 #3}
\def\np#1#2#3{{   Nucl. Phys. }{\bf #1} #2 #3 }
\def\npsup#1#2#3{{Nucl. Phys. Proc. Sup. }{\bf #1} #2 #3 }
\def\pl#1#2#3{{   Phys. Lett. }{\bf #1} #2 #3 }
\def\pr#1#2#3{{   Phys. Rev. }{\bf #1} #2 #3 }
\def\prep#1#2#3{{ Phys. Rep. }{\bf #1} #2 #3 }
\def\prl#1#2#3{{  Phys. Rev. Lett. }{\bf #1} #2 #3 }
\def\physica#1#2#3{{ Physica }{\bf #1} #2 #3 }
\def\rmp#1#2#3{{  Rev. Mod. Phys. }{\bf #1} #2 #3 }
\def\sj#1#2#3{{   Sov. J. Nucl. Phys. }{\bf #1} #2 #3 }
\def\zp#1#2#3{{   Zeit. Phys. }{\bf #1} #2 #3  }
\def\tmf#1#2#3{{  Theor. Math. Phys. }{\bf #1} #2  #3 }


\bibitem{'tHooft:1994gh}
G.~'t Hooft,
{\it Singapore, World Scientific (1994) 683 p. (Advanced series in mathematical physics, 19)}.
\bibitem{Feynman:1969ej}
R.~P.~Feynman,
Phys.\ Rev.\ Lett.\  {\bf 23} (1969) 1415.




\bibitem{thooft1}
G.~'t Hooft,
Nucl.\ Phys.\  {\bf B33} (1971) 173.

\bibitem{thooft2}
G.~'t Hooft,
Nucl.\ Phys.\  {\bf B35} (1971) 167.



\bibitem{FritzschGellmann}
H.~Fritzsch and M.~Gell-Mann,
Proc. XVI Intern. Conf. on High Energy Physics, Chicago, (1972), Vol.2. p135
\bibitem{Fritzsch:1973pi}
H.~Fritzsch, M.~Gell-Mann and H.~Leutwyler,
Phys.\ Lett.\  {\bf B47} (1973) 365.
%

\bibitem{thooftmarseille}
G. 't Hooft, Marseilles Conference on Renormalization, June 1972, 
(unpublished).

\bibitem{grosswilczekprl}
D.~J.~Gross and F.~Wilczek,
Phys.\ Rev.\ Lett.\  {\bf 30} (1973) 1343.

\bibitem{politzer}
D. Politzer, Phys.\ Rev.\ Lett.\  {\bf 30} (1973) 1346.

\bibitem{Gross:1973ju}
D.~J.~Gross and F.~Wilczek,
Phys.\ Rev.\  {\bf D8} (1973) 3633.
\bibitem{wilsonconf}
K. Wilson, Phys.\ Rev. {\bf D10} (1974) 2445. 

\bibitem{thooftconf}
G.~'t Hooft,
Nucl.\ Phys.\  {\bf B190} (1981) 455.

\bibitem{haymaker}
R.~W. Haymaker,
\prep{315}{(1999)}{153}.

\bibitem{tHooftmagmonopole}
G.~'t Hooft,
Nucl.\ Phys.\  {\bf B79} (1974) 276; 
Nucl.\ Phys.\  {\bf B105} (1976) 538. 


\bibitem{polyakov}
A.~M.~Polyakov,
\jetpl{20}{ (1974)} 1974.


\bibitem{tHooft:instantons}
G.~'t Hooft,
Phys.\ Rev.\  {\bf D14} (1976) 3432.




\bibitem{books}
QCD descibed in a number of monographs, e.g.
G. Sterman, {\it Introduction to Quantum Field Theory},
Cambridge (1993);
M.~E.~Peskin and D.~V.~Schr\"oder, {\it An Introduction to 
Qunatum Field Theory}, Addison-Wesley (1995);
 Yu.~L.~Dokshitzer \etal\  {\it Basics of 
Perturbative QCD}, Editions Frontiers (1989);
 F. Yndurain, {\it Quantum Chormodynamics}, Springer-Verlag (1983);
T. Muta, {Foundation of Quantum Chromodynamics}, 
World Scientific, Second Edition (1998);
   {\it Perturbative Quantum Chromodynamics}  Edited by  A.~H~ Mueller,
  World Scientific (1989)\,.


\bibitem{Peccei:1977hh}
R.~D.~Peccei and H.~R.~Quinn,
Phys.\ Rev.\ Lett.\  {\bf 38} (1977) 1440.


\bibitem{diagrammar}
G.~'t Hooft and M.~Veltman,
Nucl.\ Phys.\  {\bf B50} (1972) 318.


\bibitem{tHooftVeltman:dimreg}
G.~'t Hooft and M.~Veltman,
Nucl.\ Phys.\  {\bf B44} (1972) 189.



\bibitem{montvay}
   I. Montvay and G. Munster,
{\it Quantum Fields on a Lattice}.
 Cambridge (1997)\,.



\bibitem{tHooft:U(1)}
G.~'t Hooft,
Phys.\ Rev.\ Lett.\  {\bf 37} (1976) 8.


\bibitem{leutwyler1985}
J.~Gasser and H.~Leutwyler,
Nucl.\ Phys.\  {\bf B250} (1985) 465.
\bibitem{neubert}
M. Neubert, \prep{245}{(1994)}{259}


\bibitem{wu}
C.~S.~Wu, E.~Ambler, R.~W.~Hayward, D.~D.~Hoppes and R.~P.~Hudson,
Phys.\ Rev.\  {\bf 105} (1957) 1413.

\bibitem{leeyang}
T.~D.~Lee and C.~N.~Yang,
Phys.\ Rev.\  {\bf 104} (1956) 254.



\bibitem{PDGcaso}
C.~Caso {\it et al.}, Eur.\ Phys.\ J.\  {\bf C3} 1 (1998).


\bibitem{bermansirlin}
S.~M.~Berman and A. Srilin, \ap{20}{(1962)}{20}.


\bibitem{kinoshita}
T. Kinoshita, \prl{2}{(1959)}{477}\,.


\bibitem{vanRitbergen:1999fi}
T.~van Ritbergen and R.~G.~Stuart,
Phys.\ Rev.\ Lett.\  {\bf 82} (1999) 488\,.


\bibitem{glashowSM} S.~L.~Glashow, \np{B22}{(1961)}{22}.
\bibitem{weinbergSM} S. Weinberg, 
\prl{19}{(1967)}{1264}; 
 A.~Salam, Proc. 8th Nobel Symposium, (1968), ed. N. Svart;


\bibitem{Veltmanhint}
M.~Veltman,
Nucl.\ Phys.\  {\bf B7} (1968) 637.


\bibitem{Higgs:1964pj}
P.~W.~Higgs,
Phys.\ Rev.\ Lett.\  {\bf 13} (1964) 508.


\bibitem{Englert:1964et}
F.~Englert and R.~Brout,
Phys.\ Rev.\ Lett.\  {\bf 13} (1964) 321\,.



\bibitem{kamiokande}
Y.~Fukuda {\it et al.}  [Super-Kamiokande Collaboration],
Phys.\ Rev.\ Lett.\  {\bf 81} (1998) 1562.


\bibitem{GIM}
S.~L.~Glashow, J.~Iliopoulos and L.~Maiani,
Phys.\ Rev.\  {\bf D2} (1970) 1285.

\bibitem{wolfensteinpar}
L.~Wolfenstein,
Phys.\ Rev.\ Lett.\  {\bf 51} (1983) 1945.


\bibitem{buras}
A.~J.~Buras, M.~Ciuchini, E.~Franco, G.~Isidori, G.~Martinelli and L.~Silvestrini,\newline
hep-ph/0002116.

\bibitem{cecilia}
C.~Jarlskog,
Phys.\ Rev.\ Lett.\  {\bf 55} (1985) 1039.

\bibitem{kinoshitalee}
T.~Kinoshita, \jmp{3}{(1965)}{56}; 
T.~D.~Lee and M.~Nauenberg \pr{133}{(1964)}{1549}.


\bibitem{amativen13}
D.~Amati and G.~Veneziano, \np{B140}{(1978)}{54};
R~.K.~Ellis, H.~Georgi, M.~Machacek,H.~D.~Politzer and G.~G.~Gross
\np{B152}{285}{79};  
A.~V.~Efremov and A.~V.~Radyushkin, \tmf{44}{(1980)}{17};
S.~Libby and G.~Sterman, \pr{D18}{(1978)}{3252}
A.~Mueller, \pr{D18}{(1978)}{3705}

\bibitem{bodwin14}
G.~Bodwin,\pr{D31}{(1986)}{2616}; \ibid{D34}{(1986)}{3932}; 
 J.~C.~Collins, D.~E.~Soper and G.~Sterman,
\np{B261}{(1986)}{104};\ibid{B308}{(1988)}{833}

\bibitem{colsop11}
J.~C.~Collins and D.~E.~Soper, \ar{37}{(1987)}{383};
J.~C.~Collins, D.~E.~Soper and G.~Sterman,
 in Perturbative  QCD, ed. A.H. Mueller (World Scientific 1989)

\bibitem{nnloepem}
K.~G.~Chetyrkin, A.~L.~Kataev and F.~V.~Tkachov, \pl{B85}{(1979)}{277};
M.~Dine and J. Sarpinstein, \prl{43}{(1979)}{688};
W.~Celemaster and R.~Gonsalves, \prl{44}{(1980)}{560}.

\bibitem{stermanweinberg}
G.~Sterman  and S.~ Weinberg,
\prl{39}{(1977)}{1436}.

\bibitem{colsoprevs}
J.~C.~Collins and D.~E.~Soper, \ar{37}{383}{87};
J.~C.~Collins, D.~E.~Soper and G.~Sterman,
 in \lq Perturbative  QCD\rq, ed. A.H. Mueller, 
World Scientific (1989).

\bibitem{altpar}
G.~Altarelli and G.~Parisi,
Nucl.\ Phys.\  {\bf B126} (1977) 298.


\bibitem{mangano}
M.~L.~Mangano,
hep-ph/9911256.


\bibitem{abeCDFjet}
F. Abe et.~al., CDF Collab. 
Phys.\ Rev.\ Lett.\  {\bf 77} (1996) 437.

\bibitem{EKS}
S.~D.~Ellis, Z.~Kunszt and D.~E.~Soper,
Phys.\ Rev.\ Lett.\  {\bf 64} (1990) 2121.

\bibitem{ESWbook}
R.~K.~Ellis, W.~J.~Stirling and B.~R.~Webber,
{\it QCD and Collider Physics}, Cambridge University Press (1996).





\bibitem{bethke}
S.~Bethke,
hep-ex/0001023.


\bibitem{EWWG2000}
LEP and SLD  Electroweak Working Group, preprint CERN EP/2000-16.

\bibitem{Swartz:1999xv}
M.~L.~Swartz,
hep-ex/9912026 and references therein.

\bibitem{Sirlin:1999zc}
A.~Sirlin,
hep-ph/9912227  and references therein.


\bibitem{Beneke:1999fe}
M.~Beneke and A.~Signer,
Phys.\ Lett.\  {\bf B471} (1999) 233.

\bibitem{BPbook}
D. Bardin and G. Passarino,
 {\it The Standard Model in the Making}, Clarendon Press,
 Oxford (1999).




\bibitem{altbar}
G.~Altarelli and R.~Barbieri,
Phys.\ Lett.\  {\bf B253} (1991) 161;
G. Altarelli, R. Barbieri and F. Caravaglios,
\np{B405}{3}{(1993)} \,.


\bibitem{steinhauser}
M. Steinhauser,
Phys.\ Lett.\  {\bf B429} (1998) 158.


\bibitem{Jegerlehner:1999hg}
F.~Jegerlehner,
hep-ph/9901386.


\bibitem{veltman1977}
M.~Veltman,
Nucl.\ Phys.\  {\bf B123} (1977) 89.



\bibitem{barbieri-efftop}
R.~Barbieri, M.~Beccaria, P.~Ciafaloni, G.~Curci and A.~Vicere,\newline
Nucl.\ Phys.\  {\bf B409} (1993) 105.

\bibitem{vanderbij}
J.~J.~van der Bij and F.~Hoogeveen,
Nucl.\ Phys.\  {\bf B283} (1987) 477.




\bibitem{Degrassi:1999jd}
G.~Degrassi and P.~Gambino,
hep-ph/9905472.





\bibitem{D'Agostini:2000ws}
G.~D'Agostini and G.~Degrassi,
hep-ph/0001269.

\bibitem{ApBe80}
 T. Appelquist and C. Bernard Phys.\ Rev.\  {\bf D22}  (1980) 200. 

\bibitem{longhitano:81}
A. Longhitano, Phys.\ Rev.\  {\bf D22} (1980)  1166;\  
\ Nucl.\ Phys. {\bf B188}(1981) 118. 

\bibitem{pestak}
M. Peskin and T. Takeuchi, \prl{65}{(1990)}{2963}.


\bibitem{Bagger:1999te}
J.~A.~Bagger, A.~F.~Falk and M.~Swartz, hep-ph/9908327.



\bibitem{veltman}
M. Veltman,  Act.\ Phys.\ Pol.\ {\bf B8} (1977) 475.


\bibitem{bardin}
D. Bardin, \etal\ Comp Phys. Comm. {\bf 104} (1997) 161.


\bibitem{LeeQuigg}
B.~W.~Lee, C.~Quigg and H.~B.~Thacker,
Phys.\ Rev.\  {\bf D16} (1977) 1519.


\bibitem{Frohlich}
J.~Fr\"ohlich,
Nucl.\ Phys.\  {\bf B200} (1982) 281; 
M.~Aizenman,
Phys.\ Rev.\ Lett.\  {\bf 47} (1981) 1.

\bibitem{kuti} See for example
J.~Kuti, L.~Lin and Y.~Shen,
Phys.\ Rev.\ Lett.\  {\bf 61} (1988) 678.

\bibitem{altisi}
G.~Altarelli and G.~Isidori,
Phys.\ Lett.\  {\bf B337} (1994) 141.

\bibitem{quiros}
J.~A.~Casas, J.~R.~Espinosa and M.~Quiros,
Phys.\ Lett.\  {\bf B382} (1996) 374

\bibitem{hambye}
T.~Hambye and K.~Riesselmann,
hep-ph/9708416.

\bibitem{guide} J.~F.~Gunion, H.~E.~Haber, G.~L.~Kane and S.~Dawson, 
                {\it ``The Higgs Hunter Guide''} 
                Addison-Wesley, Reading MA, (1990).




\bibitem{LHC}
 Proceedings of 
the ``{\it Large Hadron Collider Workshop}'', Aachen, 4-9 October    1990,
 eds. G.~Jarlskog and D.~Rein, Report CERN 90-10, ECFA 90-133, Geneva, 1990.  

\bibitem{spirahgludec} M. Spira, CERN-TH/96-285 and references therein

\bibitem{corrreview} B.~Kniehl, DESY Report No.~93-069; 
        M.~Carena, P.M.~Zerwas {\it et al.},
                     in ``Physics at LEP2", eds. G.~Altarelli et.al., 
                     CERN Report 96-01, Vol.1, p.351 (1996). 

\bibitem{oldpaper} Z.~Kunszt and W.J.~Stirling, in Ref.~\cite{LHC}.
\bibitem{MWJSZK}
Z.~Kunszt, S.~Moretti and W.~J.~Stirling,
Z.\ Phys.\  {\bf C74} (1997) 479\,.



\bibitem{xggh} H.~Georgi, S.L.~Glashow, M.E.~Macahek and D.V.~Nanopoulos,
               \prl{40}{(1978)}{692}.
               
\bibitem{xvvh} R.~N.~Cahn and S.~Dawson, \pl{B136}{(1984)}{196}.
   
\bibitem{xvh} S.~L.~Glashow, D.~V.~Nanopoulos and A.~Yildiz,
 \pr{D18}{(1978)}{1724}; 
R.~Kleiss, Z.~Kunszt and W.~J.~Stirling,
Phys.\ Lett.\  {\bf B253} (1991) 269; 
  Z.~Kunszt, Z.~Trocsanyi and W.J.~Stirling, \pl{B271}{(1991)}{247}.

\bibitem{xtth} Z.~Kunszt, \np{B247}{(1984)}{339}; 
               J.~F.~Gunion, \pl{B253}{(1991)}{269}; 
               W.~J.~Marciano and F.~E.~Paige, \prl{66}{(1991)}{2433}; 
               J.F.~Gunion, H.~E.~Haber, F.~E.~Paige, W.-K. Tung and
               S.~S.~D. Willenbrock, \np{B294}{(1987)}{621}; 
               D.~A.~Dicus and  S.~S.~D. Willenbrock, 
\pr{D39}{(1989)}{751}.        

\bibitem{newKfacgg} D.~Graudenz, M.~Spira and P.M.~Zerwas, 
\prl{70}{( 1993)}{1372}.


\bibitem{Kfacgg2} S.~Dawson, \np {B359}{(1991)}{283}.

\bibitem{Kfacgg3} S.~Dawson and R.~P.~Kauffman, \pr{ D49}{(1993)}{2298}.


\bibitem{DjouadiRev} A.~Djouadi, {\it Int. J. Mod. Phys.} {\bf A10}
                     (1995) 1; 


\bibitem{sdgz}
M.~Spira, A.~Djouadi, D.~Graudenz and P.~M.~Zerwas, \np {B453}{(1995)}{17}.

\bibitem{ATLAS} ATLAS Technical Proposal, 
CERN/LHC/94-43 LHCC/P2 (1994).

\bibitem{CMS} CMS Technical Proposal, CERN/LHC/94-43 LHCC/P1 (1994).
                       
\bibitem{richter}
D. Froidevaux, 
 Elzbieta Richter-Was,    CERN-TH-96-111 ( 1996).


\bibitem{dittdrei}
M.~Dittmar and H.~Dreiner,
Phys.\ Rev.\  {\bf D55} (1997) 167\,.

\bibitem{atlas-TDR}
ATLAS Technical Design Report, Volume II, CERN (1999). 

\end{thebibliography}
\end{document}